\documentclass[11pt, letterpaper]{berkeley}

\usepackage{amsmath}
\usepackage{hyperref}
\usepackage{url}
\usepackage{graphicx}
\usepackage{booktabs}
\usepackage{wrapfig}
\usepackage{multirow}
\usepackage{lipsum}
\usepackage{subcaption}
\usepackage{amssymb}
\usepackage{tikz} 
\usepackage{xcolor}
\usepackage{multicol}
\usepackage{natbib}
\usepackage{listings}
\usepackage{fancyvrb}
\usepackage{fvextra}
\usepackage{tcolorbox}
\usepackage{xcolor}
\usepackage{hyperref}
\usepackage{adjustbox}
\usepackage{twemojis}
\usepackage{pgfplots}
\usepackage{animate}
\usepgfplotslibrary{groupplots}
\usetikzlibrary{patterns}

\definecolor{brownishyellow}{RGB}{245, 194, 66}
\definecolor{band}{RGB}{96, 36, 0}
\definecolor{comp}{RGB}{158, 80, 33}
\definecolor{mem}{RGB}{62, 97, 123}
\definecolor{tri}{RGB}{150, 24, 32}
\newtcolorbox{solutionbox}{
  colframe=black,
  colback=gray!10,
  boxrule=1pt,
  arc=0pt,
  title=,
  fonttitle=\bfseries
}

\newenvironment{sol}
  {\begin{solutionbox}}
  {\end{solutionbox}}

\newcommand{\BeginSol}{\begin{sol}}
\newcommand{\EndSol}{\end{sol}}

\definecolor{mylightgray}{RGB}{240,240,240}

\usepackage{tcolorbox}
\tcbuselibrary{skins}
\newtcolorbox{text_full}[1]{
    enhanced,
    left=4mm,
    right=4mm,
    top=2mm,
    bottom=2mm,
    boxsep=0mm,
    rounded corners,
    title=#1,
    fontupper=\footnotesize\linespread{0.9}\fontfamily{lmr}\selectfont,
    }
\newtcolorbox{text_half}[1]{
    enhanced,
    left=2mm,
    right=2mm,
    top=2mm,
    bottom=2mm,
    boxsep=0mm,
    rounded corners,
    title=#1,
    width=0.475\textwidth,
    fontupper=\footnotesize\linespread{0.9}\fontfamily{lmr}\selectfont,
    }

\newcommand\correspondencefootnote{%
  \renewcommand{\thefootnote}{}%
  \footnotetext{}%
  \renewcommand{\thefootnote}{\arabic{footnote}}%
}

\newcommand{\zgfrowlabel}[1]{%
    \llap{%
        \raisebox{0pt}[0.1875\linewidth][0pt]{%
            \vbox to 0.1875\linewidth{\vfil\hbox{\rotatebox{90}{\scriptsize\texttt{#1}}}\vfil}%
        }\hspace{2pt}%
    }%
}
\title{Computation-Bandwidth-Memory Trade-offs: A Unified Paradigm for AI Infrastructure}
\author{Yuankai Fan}
\author{Qizhen Weng}
\author{Xuelong Li\footnote{Corresponding author: Xuelong Li (\url{xuelong_li@ieee.org})\\\makebox[\linewidth][l]{\hspace{2em}Date: December 30, 2025}}}
\affil{Institute of Artificial Intelligence (TeleAI), China Telecom}

\begin{document}
\correspondencefootnote

% \blfootnote{Authors are ordered randomly. Correspondence to \href{mailto:team@synthlabs.ai}{team@synthlabs.ai}.}

\begin{abstract}
 % Finding the balance among the three that maximizes efficiency is difficult. 
 Large-scale artificial intelligence (AI) models are fundamentally transforming industries and redefining the paradigm of human–machine collaboration. While the technological revolution signals a new era of machine intelligence, the continued scaling of these models has exposed significant limitations in contemporary hardware architectures, manifesting as constraints on computational efficiency, interconnection bandwidth, and memory capacity. These three dimensions are inseparably intertwined, such that advances along any single axis often exacerbate bottlenecks in the others, rendering isolated optimizations increasingly ineffective. Achieving an optimal balance among them to maximize system efficiency therefore remains a central challenge in the design of scalable AI systems.
 To address this challenge, we introduce \textbf{Computation-Bandwidth-Memory Trade-offs}, termed the \textbf{AI Trinity}, a unified paradigm that positions \textit{computation}, \textit{bandwidth}, and \textit{memory} as coequal pillars for next-generation AI infrastructure. At its core, AI Trinity enables a dynamic flow of resources across these pillars, transcending single-resource bottlenecks and adapting to diverse scenarios, thereby optimizing overall system performance to its maximum potential.
 Within this framework, AI Trinity identifies three fundamental trade-offs: (1) \textbf{More Computation$\rightarrow$Less Bandwidth}, wherein computational resources are exploited to reduce data transmission under limited bandwidth conditions, (2) \textbf{More Bandwidth$\rightarrow$Less Memory}, which exploits abundant communication capacity to populate or refresh memory when local storage resources are constrained, and (3) \textbf{More Memory$\rightarrow$Less Computation}, whereby storage capacity are utilized to mitigate redundant computation when computational costs are prohibitive.
 We illustrate the effectiveness of AI Trinity through representative system designs spanning edge–cloud communication, large-scale distributed training, and model inference. The innovations embodied in AI Trinity advance a new paradigm for scalable AI infrastructure, providing both a conceptual foundation and practical guidance for a broad range of application scenarios. 
 % including but not limited to embodied AI, world models, and underwater vehicles.
\end{abstract}

\maketitle

\begin{flushright}
\textit{The whole is greater than the sum of its parts. \\ -Aristotle}
\end{flushright}

\tableofcontents

\clearpage

\section{Introduction}
Artificial intelligence (AI) has fundamentally transformed human–computer interaction by enabling machines to perform increasingly sophisticated cognitive tasks that traditionally require human intelligence.
% The ``\textit{compression is intelligence}'' since the 90's, viewing the ability to learn efficient representations of data as a core component of intelligent behavior~\citep{hernandez1998, Matthew99}, has gained renewed attention under the modern large-scale models. %% qizhen: cannot make it
A major milestone in this transformation was the release of ChatGPT in November 2022, a large-scale language model (LLM) developed by OpenAI, demonstrating how AI technologies are reshaping interactions and becoming deeply integrated into both daily life and various industries.
% The rapid progress of LLMs has also revived long-standing theoretical principles, often summarized as ``\textit{compression is intelligence}'', which views the ability to learn efficient representations of data as a core component of intelligent behavior~\citep{hernandez1998,Matthew99}. This perspective has gained renewed attention as modern large-scale models increasingly demonstrate strong empirical performance across a wide range of tasks. %% qizhen: seems to me the above sentence deviating from the main line.
By 2024, a diverse array of advanced AI models\textemdash including LLMs such as GPT-4o~\citep{GPT-4o-24}, LLaMa-3~\citep{Llama3-24} and DeepSeek-V3~\citep{DeepSeek24}, as well as generative models like DALL·E 3~\citep{Dalle3_2023} and Sora~\citep{openai2024sora}\textemdash have further advanced the field, progressing closer to the realization of Artificial General Intelligence.

% Empirical scaling laws~\citep{Scaling20} show that model performance improves predictably with larger model sizes, more training data, and greater computational resources, driving the rapid expansion of modern AI systems."
% Despite these remarkable achievements, the rapid evolution of AI models underscores the central role of systematic scaling in driving performance gains. %% qizhen: highlight the scaling
{The scaling of AI infrastructure}, despite its high cost, has emerged as one of the central roles in driving these remarkable achievements~\citep{sutton2019bitter}.
Empirical scaling laws~\citep{Scaling20} show that model performance rises predictably with larger model size, more training data, and greater computational resources, thereby fueling the unprecedented growth of modern AI systems.
Consequently, leading AI companies, including OpenAI, Google, and Meta, are channeling investments ranging from tens to hundreds of billions of dollars into expanding their AI infrastructure, so as to secure the computational power required for frontier models.

This scaling trajectory, however, is gradually constrained by the physical and economic boundaries associated with the plateauing of Moore's Law~\citep{Moore65}, presenting critical challenges for current hardware architectures~\citep{Insights-v3-25}.
Due to the limited high-bandwidth memory on GPUs, training a trillion-parameter model requires roughly on the order of $1,024$ GPUs interconnected via multiple $400\text{ Gbps}$ RDMA-capable network interface cards~\citep{ZeRO20}.
During the inference of trillion-parameter LLM, serving batches of $128$ sequences (each with $2,048$ tokens) demands about $1.2\text{ TB}$ of memory to only store the key–value cache and around $350\text{ GB}$ for the model parameters in 16-bit precision~\citep{Tianyi24} \textemdash Both training and inference of large-scale models place heavy pressure computation power, communication bandwidth, and memory capacity.
% Similarly, during inference, a moderately sized $8$B-parameter LLM serving batches of tens of requests (each comprising $10,000$ tokens) requires approximately $12\text{ GB}$ of memory to store the key–value (KV) cache and $16\text{ GB}$ for the model parameters using the widely adopted 16-bit precision~\citep{Jenga25}.

% In practice, resource provisioning in large-scale AI systems is often imbalanced, resulting in situations where certain resources become critical bottlenecks while others remain insufficiently exploited.
Suboptimal resource allocation in AI systems further intensifies the economic waste and environmental impact of the inflating, power-intensive AI infrastructure.
For example, an edge–cloud AI system may saturate network bandwidth due to excessive data offloading, while increasingly powerful computational resources in edge devices remain underexploited~\citep{Neurosurgeon17}. Conversely, an edge computing system exhibits limited computational capacity, whereas storage resources remain comparatively abundant~\citep{fog23}. Although prior studies have sought to improve system performance by alleviating single-resource bottlenecks under specific scenarios, such narrowly scoped optimizations lack adaptability across diverse operating conditions. Thus, achieving an optimal balance among computation, bandwidth, and memory to maximize overall system efficiency requires fundamental advances in AI system design, underpinned by a widely accepted set of standardized frameworks and protocols.

\begin{figure}[h]
    \centering
    \includegraphics[width=\linewidth]{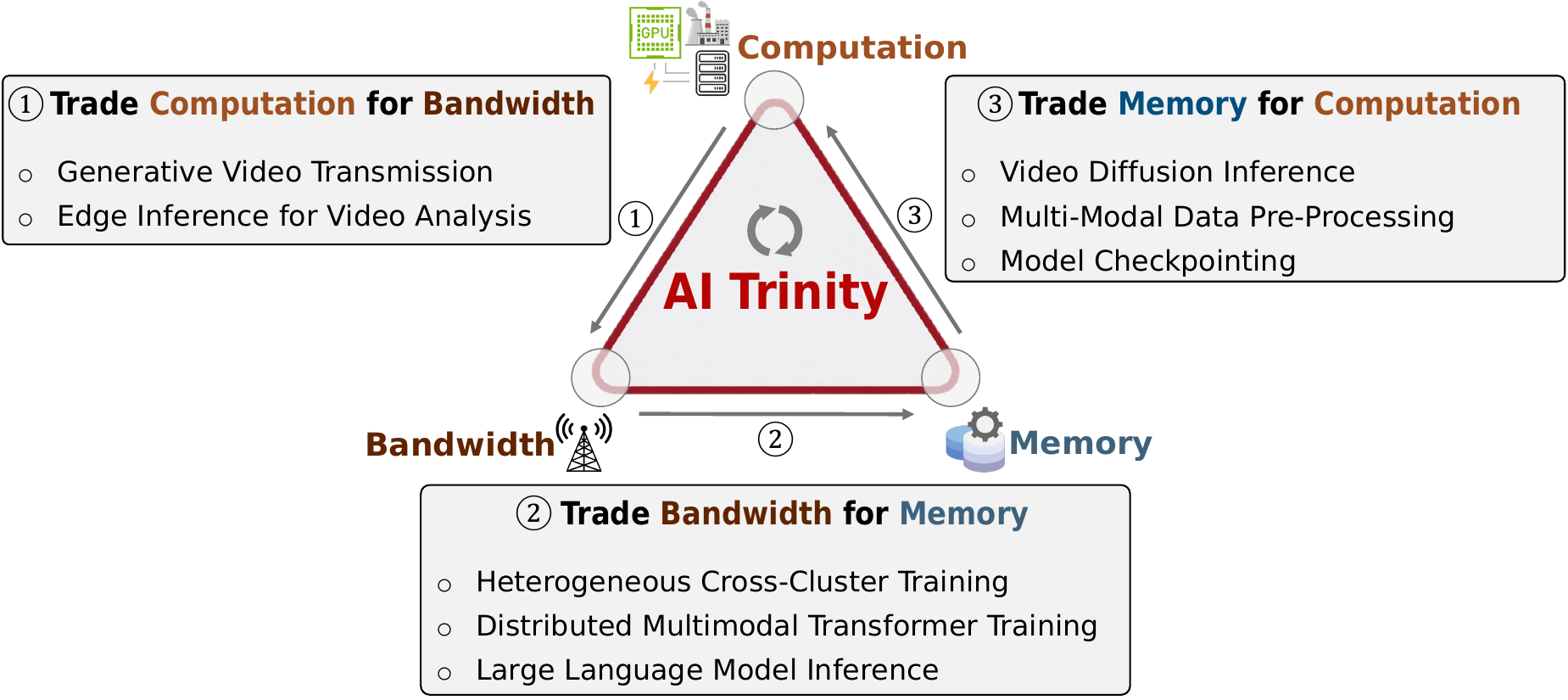}
    \caption{An overview of Computation-Bandwidth-Memory Trade-offs, \textit{i.e.}, the \textcolor{tri}{\textbf{AI Trinity}} framework.}
    \label{fig:overview}
\end{figure}

In this paper, we introduce \textbf{Computation-Bandwidth-Memory Trade-offs}, termed the \textbf{AI Trinity}, a unified design principle for AI systems. At its core, \textbf{AI Trinity aims to achieve system-level performance optimality by dynamically balancing computation, bandwidth, and memory through scenario-aware resource exchanges}. Rather than optimizing a single resource in isolation, AI Trinity explicitly leverages resource substitution, \textit{i.e.}, trading computation, bandwidth, and memory according to workload characteristics and system constraints, to adapt to heterogeneous edge–device–cloud environments~\citep{ShaoZ20, Shao24, AI-Flow25, chen2025GVC}. As illustrated in~\autoref{fig:overview}, AI Trinity formulates a closed-loop set of resource exchange pathways and identifies three fundamental trade-offs as follows:

\begin{enumerate}
    \item[\textcircled{1}] \textit{More Computation for Less Bandwidth} \textbf{(\textcolor{comp}{Computation}$\rightarrow$\textcolor{band}{Bandwidth})}. This trade-off leverages additional computational resources to reduce communication overhead by performing partial data processing locally. This mechanism is particularly beneficial in bandwidth-constrained scenarios, where unprocessed data offloading would otherwise dominate latency and energy consumption. By shifting computation closer to the data source, AI Trinity mitigates network bottlenecks while maintaining end-to-end system throughput.
    % Formally, let $D$ denote data volume and $C$ denote available computation; by increasing $C$ to pre-process or compress $D$ before transmission, the effective bandwidth demand $B$ is reduced ($B \propto f(D,C)$).
    
    \item[\textcircled{2}] \textit{More Bandwidth for Less Memory} \textbf{(\textcolor{band}{Bandwidth}$\rightarrow$\textcolor{mem}{Memory})}. This trade-off exploits higher communication capacity to alleviate memory constraints by dynamically offloading data or model parameters to remote storage. Such a strategy is critical for memory-limited scenarios, enabling the system to maintain operational correctness while supporting larger workloads.
    % Denoting local memory as $M$ and bandwidth as $B$, increasing $B$ allows effective memory footprint $M_\text{eff}$ to be reduced ($M_\text{eff} \propto f(M,B)$) without impacting computational correctness.
    
    \item[\textcircled{3}] \textit{More Memory for Less Computation} \textbf{(\textcolor{mem}{Memory}$\rightarrow$\textcolor{comp}{Computation})}. This trade-off uses excess memory resources to store intermediate computations or precomputed features, reducing redundant computation. This approach is particularly effective in compute-intensive scenarios, improving execution efficiency and lowering massive computational cost.
    % Let $C_\text{req}$ denote required computation; by leveraging additional memory $M$ to store reusable artifacts, $C_\text{req}$ can be reduced ($C_\text{req} \propto f(M)$).
\end{enumerate}

% \begin{table}[t]
% \centering
% \caption{Scenario-aware resource bottlenecks and corresponding AI Trinity trade-offs.}
% \label{tab:trinity_scenarios}
% \begin{tabular}{l l l}
%     \toprule
%     \textbf{Scenario} & \textbf{Dominant Bottleneck} & \textbf{AI Trinity Trade-off} \\
%     \midrule
%     Edge--cloud offloading with limited uplink &
%     Network bandwidth &
%     Computation $\rightarrow$ Bandwidth \\
%     \addlinespace
    
%     Memory-constrained edge inference &
%     On-device memory &
%     Bandwidth $\rightarrow$ Memory \\
%     \addlinespace
    
%     Compute-limited edge devices with repetitive workloads &
%     Computation &
%     Memory $\rightarrow$ Computation \\
%     \addlinespace
    
%     Large-scale model training with heterogeneous nodes &
%     Imbalanced multi-resource usage &
%     Adaptive combination of Trinity trade-offs \\
%     \bottomrule
% \end{tabular}
% \end{table}

We demonstrate the applicability and generality of AI Trinity through representative AI system designs spanning training, inference, and the full AI development lifecycle. These examples include intelligent communication systems that reduce communication cost by introducing additional computation, distributed training paradigms that exploit communication to effectively extend memory capacity, and inference serving systems that reuse cached representations to eliminate redundant computation. Collectively, they illustrate how AI Trinity translates the principle of resource balancing into concrete design choices, enabling near-optimal system performance by mitigating single-resource bottlenecks in heterogeneous environments. \textbf{By providing a systematic lens for reasoning about performance bottlenecks, AI Trinity empowers practitioners to rethink system design decisions beyond naive resource scaling}.

The remainder of the paper is organized as follows: Section~\ref{sec:2} provides a brief review of related work. Section~\ref{sec:3} introduces the ``computation-for-bandwidth'' pathway and presents two scenarios illustrating this trade-off. Section~\ref{sec:4} explores the ``bandwidth-for-memory'' pathway, while Section~\ref{sec:5} details the ``memory-for-computation'' pathway and its implementation in specific scenarios. Finally, Section~\ref{sec:6} discusses future directions for expanding its scope,  and Section~\ref{sec:7} concludes the paper.

\section{Related Work}\label{sec:2}
The design of efficient AI systems has long been shaped by the advent of big data and the increasing scale of computational workloads. Early work in distributed computing formalized fundamental trade-offs between storage, computation, and communication, showing that improving one resource often comes at the expense of others~\citep{Qifa18}. These insights laid the foundation for modern AI system design, where holistic, resource-aware strategies are critical to achieving scalable and efficient performance.

Recent advancements in deep neural networks (DNNs) has further explored explicit resource substitution mechanisms to improve system efficiency. For example, SmartExchange~\citep{SmartExchange20} investigates trading higher-cost memory storage or access for lower-cost computation, demonstrating that flexible resource exchange can significantly reduce overall system cost under dynamic workloads. With the recent success of LLMs and the growing demand for low-latency inference, several works have focused on efficiently managing computation and memory resources. For instance, Mooncake~\citep{Mooncake25}, the serving platform for Kimi~\citep{kimi25}\textemdash an LLM service provided by Moonshot AI\textemdash presents a KV-cache-centric design for serving LLM-based chatbots, where excess memory is leveraged to store precomputed representations, reducing redundant computation during inference. While these approaches highlight the practical benefits of computation–memory trade-offs in specific scenarios, they only partially address the broader resource interactions. In contrast, AI Trinity provides a unified framework that systematically balances computation, memory, and bandwidth to achieve near-optimal system-level performance across heterogeneous environments.

On the other hand, in edge and cloud computing, resource imbalance is especially pronounced due to heterogeneous capabilities across devices and network constraints. Previous approaches have proposed to address isolated resource bottlenecks, such as offloading computation to alleviate bandwidth limitations~\citep{Neurosurgeon17}, or using distributed memory to support large-scale model training~\citep{fog23}. However, these approaches often lack generality, as they do not simultaneously consider multiple resource trade-offs. AI Trinity builds upon these foundations by providing a unified framework that explicitly formalizes the substitution between computation, memory, and bandwidth resources, enabling adaptive optimization across diverse scenarios in modern AI systems.

\section{Trading Computation for Bandwidth}\label{sec:3}
 Recent advancements in large-scale models have driven substantial progress across a wide range of domains, showcasing remarkable capabilities and unprecedented potential. These advances have enabled their adoption in diverse application areas, including embodied robotics~\citep{Embodied-AI25}, augmented reality, and autonomous driving~\citep{DriveGPT24}. Meanwhile, the rapid proliferation of edge devices, such as smartphones, smart wearables, and IoT sensors, has made sensory data increasingly accessible. In the era of ubiquitous connectivity, leveraging communication networks to distribute intelligence has emerged as a transformative paradigm, enabling AI-powered services to be deployed and accessed efficiently at the network edge~\citep{Shao24, AI-Flow25}.

 However, enabling intelligence at scale across distributed edge environments inevitably exposes a fundamental trade-off between computation and communication~\citep{ShaoZ20}. While offloading computation to edge or cloud servers can alleviate the limited processing capabilities of end devices, it simultaneously increases reliance on communication networks to transmit intermediate or collaborative data. As a result, distributed and collaborative inference incurs substantial communication overhead because high-dimensional data must be exchanged frequently, heavily straining network bandwidth and creating significant latency bottlenecks.

 % For example, end devices must transmit intermediate activation features to edge servers for subsequent processing for edge inference, often generating tens to hundreds of megabytes of data per inference step. The resulting cumulative data transfers can overwhelm network infrastructure, leading to increased latency and reduced throughput. Similarly, in LLM-based multi-agent systems, the iterative synchronization of reasoning traces among agents further amplifies communication overhead, thereby limiting system scalability.

 \begin{figure}[h]
    % \vspace{-4mm}
    \includegraphics[width=\linewidth]{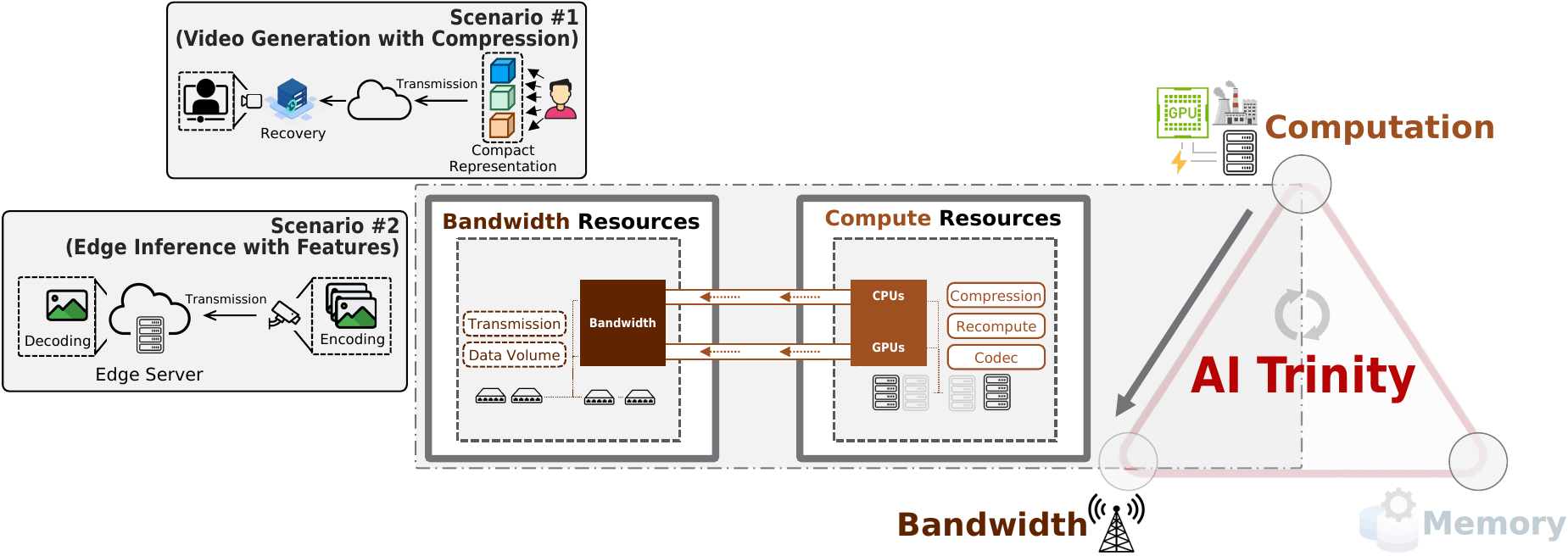}
    \caption{More \textbf{\textcolor{comp}{Computation}} for less \textbf{\textcolor{band}{Bandwidth}} trade-off.}
    \label{fig:comp-band}
\end{figure}

 For instance, in edge inference, end devices are required to send intermediate activation features to edge servers for continued computation, resulting producing up to several hundred megabytes of data generated per inference step. Such repeated transfers can saturate network resources, resulting in higher latency and lower throughput. Likewise, in modern LLM-based multi-agent systems, the repeated synchronization across agents further exacerbates communication overhead, further amplifies communication overhead, thereby limiting system scalability.

 In summary, ubiquitous AI faces a dual-bottleneck challenge arising from constrained hardware resources and limited communication capacity. \textbf{This challenge highlights the necessity of communication-efficient system designs that carefully balance computation and communication to reduce bandwidth consumption while maintaining system accuracy and service quality}.

\autoref{fig:comp-band} illustrates this computation–communication trade-off through two representative scenarios. In the next two subsections, we examine these scenarios in detail. Specifically, we show how transmission bandwidth can be effectively reduced in edge–cloud AI systems, and how advanced feature representations can further alleviate bandwidth consumption in edge inference settings.

\subsection{Generative Video Transmission with Compression}\label{sec:2.1}
\subsubsection{Background}
Video transmission is among the most bandwidth-intensive workloads in modern communication systems. With the widespread deployment of high-resolution cameras and video sensors on edge devices (\textit{e.g.}, surveillance cameras, smartphones, and autonomous vehicles), raw video streams are increasingly generated at high frame rates and resolutions. The resulting data volume grows rapidly with spatial resolution, frame rate, and color depth, producing raw bitrates that far exceed the capacity of most practical communication links.

\noindent\textbf{Challenges}. In real-world systems, these video streams are often transmitted over bandwidth-constrained networks for next-stage processing.  Unfortunately, transmitting raw video data in such settings is impractical: the sheer volume of data would quickly saturate available bandwidth, leading to network congestion and increased latency. These limitations are particularly pronounced in edge and mobile environments, where bandwidth availability is inherently limited.
% Typical scenarios include live video streaming to cloud servers for content delivery and offloading video frames to edge servers for perception and decision-making tasks.

\textit{Video compression} is therefore a fundamental enabling technology for scalable video transmission. The core idea of compression is to exploit redundancy and structure in video data to reduce the number of bits required for representation while preserving perceptual or task-relevant information. However, both traditional codecs~\citep{VVC20,AV1} and recent neural video compression approaches~\citep{li2021deep, li2023neural, li2024neural} often struggle at ultra-low bitrates, producing videos with poor perceptual quality manifested as blurring, blocking artifacts, and color distortions. \textbf{These limitations motivate approaches that trade transmission cost for computation, shifting the burden from bit allocation to inference while preserving semantic richness}.

% \noindent\textbf{Proposed Method.}
\noindent\textbf{Solutions.}
To overcome these limitations, related works introduce a diffusion-based video compression framework that reformulates video compression as a conditional generation task~\citep{Li2024PiNoise, Fangqiu25}. In particular, the framework designs compact representations tailored to segmentation sequences (\textit{i.e.}, geometric scaffolding that preserves object boundaries and spatial relationships across frames), human motion data (\textit{i.e.}, articulated body dynamics that capture temporally coherent motion patterns), and optical flow (\textit{i.e.}, encodings of dense pixel-wise displacement vectors between consecutive frames), all of which can be efficiently transmitted at very low bitrates. These representations preserve the essential semantic and structural cues of the original video, enabling the generative model to reconstruct visually coherent, semantically rich video at minimal transmission cost.
% Instead of transmitting densely encoded pixel data, the sender transmits sparse yet informative signals, and a generative diffusion model synthesizes the corresponding video conditioned on these signals. 

\subsubsection{Exploring the Potential of the Trade-offs}
This section aims to formalize how increased computational effort at the receiver can compensate for reduced transmission bandwidth, while still preserving perceptual quality and semantic fidelity. We consider a video transmission system comprising a sender (edge device) and a receiver (edge server or cloud), connected via a bandwidth-limited communication channel. 

Let $V$ denote the original video sequence, and let $R$ be the transmitted representation produced by an encoder. The overall system can be abstracted as:
$$
R = \mathcal{E}(V), \quad \hat{V} = \mathcal{D}(R),
$$
where $\mathcal{E}(\cdot)$ is the encoding and transmission process, and $\mathcal{D}(\cdot)$ is the reconstruction process.

We characterize each transmission strategy by three key quantities: (1) transmission bandwidth cost $B(R)$, measured in bits per pixel (BPP), (2) decoding computation cost $C(\mathcal{D})$, measured in FLOPs, and (3) reconstruction quality $Q(\hat{V}, V)$, measured in task-oriented metrics.

\noindent\textbf{Bandwidth-dominated Regime.}
In traditional video compression pipelines, such as H.264 or neural codecs, the decoder is intentionally lightweight to enable real-time playback on on devices with limited resources. As a result, the achievable quality is primarily limited by the bitrate:
$$
Q \approx f(B), \quad \text{with } C(\mathcal{D}) \text{ constrained to be small}.
$$
% At ultra-low bitrates, aggressive quantization and motion compensation lead to irreversible information loss, resulting in blocking artifacts, temporal flicker, or over-smoothed textures.

\noindent\textbf{Computation-amplified Reconstruction}. In contrast, diffusion-based video compression operates in a fundamentally different regime. Instead of reconstructing pixels deterministically from dense bitstreams, the decoder leverages a powerful generative model to synthesize plausible video content conditioned on sparse signals:
$$
\hat{V} \sim p_\theta(V \mid R),
$$
where $p_\theta$ is parameterized by a diffusion model with substantial computational capacity.

In this setting, reconstruction quality depends jointly on bandwidth and computation:
$$
Q \approx g(B, C),
$$
where increasing decoding computation $C$ can partially or even largely offset reductions in bandwidth $B$. Intuitively, sparse representations provide high-level structural constraints, while the generative model fills in missing details through learned priors.

\noindent\textbf{Trade-offs}.
This interaction induces a natural trade-off frontier. For a fixed quality target $Q_0$, the minimal required bandwidth can be expressed as:
$$
B^*(Q_0, C) = \min_{R} B(R)
\quad \text{s.t.} \quad Q(\hat{V}, V) \ge Q_0, C(\mathcal{D}) \le C.
$$

As $C$ increases, $B^*(Q_0, C)$ decreases monotonically, revealing that additional computation enables more aggressive compression. In the extreme case, when $C$ is sufficiently large, the transmitted representation may shrink to only a few key frames or sparse conditioning signals, with most visual content generated at the receiver.

\begin{wrapfigure}{r}{0.55\textwidth}
    \includegraphics[width=\linewidth]{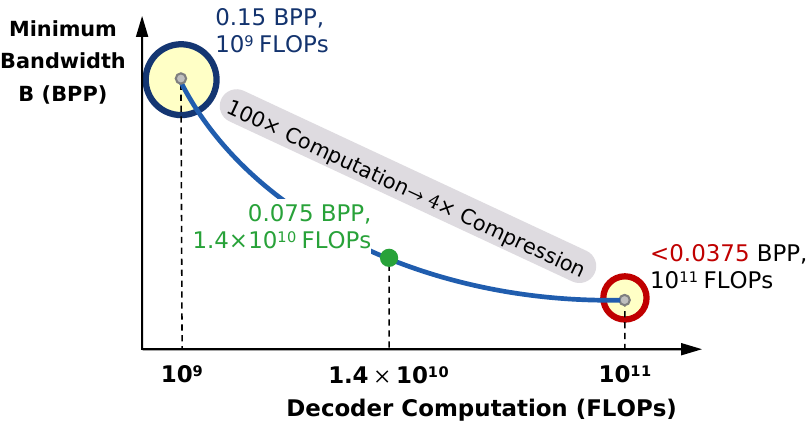}
    \caption{Computation vs. bandwidth example.}
    \label{fig:bpp}
\end{wrapfigure}
\noindent\textbf{An illustrated example}. Assume a target perceptual quality $Q_0$ is fixed. Using a conventional H.264 approach transmits a dense representation at $B = 0.15$ BPP, decoded with a lightweight model costing $10^9$ FLOPs. By increasing decoder computation to $C \approx 1.4\times10^{10}$ FLOPs, the same quality $Q_0$ can be maintained while reducing the transmission to $0.075$ BPP. Further increasing computation to $10^{11}$ FLOPs allows $B$ to drop below $0.0375$ BPP. As shown in~\autoref{fig:bpp}, such a trade-off can be interpreted as roughly ``$100\times$ more computation $\rightarrow$ $4\times$ compression'' under this scenario.

\subsubsection{Preliminary Results}\label{sec:2.1.3}
% \begin{figure}[h]
%     \centering
%     \includegraphics[width=\textwidth]{figures/output.png}
%     \vspace{-8mm}
%     \caption{R-D (perception) performance on different datasets. \textbf{Top}: FVD (↓); \textbf{Bottom}: LPIPS (↓).}
%     \label{fig:sota}
% \end{figure}

Existing works have demonstrated the promise of video compression in the ultra-low-bitrate regime. Readers may find more details of generative video compression in these works~\citep{Fangqiu25, chen2025GVC}. Below, we briefly highlight the key observations.

\noindent\textbf{Tasks \& Datasets}. The evaluation primarily focuses on the video compression task using public benchmarks: including HEVC Class B and C~\citep{HEVC_class_B_and_C}, UVG~\citep{Mercat_2020_UVG}, as well as MCL-JCV~\citep{Wang_2016_MCL_JCV}, following~\citep{wan2024m3cvccontrollablevideocompression} and T-GVC~\citep{Wang_2025_T_GVC}.

\noindent\textbf{Evaluation Metrics}. Two metrics, namely Frechet Video Distance (FVD) as well as Learned Perceptual Image Patch Similarity (LPIPS), are used for evaluation. Lower FVD and LPIPS at the same bitrate represent better performance.

\noindent \textbf{Baselines}. Following are the baseline methods for comparison: (1) traditional video compression codecs, H.264~\citep{Wiegand_2003_H264}/H.265~\citep{Sullivan_2012_H265}/H.266~\citep{Bross_2021_H266}; (2) neural codecs, {DCVC-RT}~\citep{jia2025towards}, {DCVC-FM}~\citep{li2024neural}, and {DCVC-DC}~\citep{li2023neural}; (3) diffusion-based video compression methods ({T-GVC}).

\noindent\textbf{Compression Settings}. The work employs four compression levels ($0$-$3$), each corresponding to a distinct feature-extraction strategy for segmentation sequences, human motion, and optical flow, yielding different BPP values.

% \begin{table}[h]
%     \centering
%     \begin{tabular}{|c|c|c|c|c|}
%         \hline
%         \multirow{2}{*}{Level} & \multicolumn{3}{c|}{Features} & \multirow{2}{*}{BPP} \\
%         \cline{2-4}
%         & Segmentation Sequence & Human Motion & Optical Flow & \\
%         \cline{1-5}
%         0 & N/A & N/A & N/A & $0.0024$ \\
%         1 & $N = 10$ & $\xi = 1 / 5$ & $l = 128$ & $0.0066$ \\
%         2 & $N = 20$ & $\xi = 1 / 8$ & $l = 96$ & $0.0099$ \\
%         3 & $N = 30$ & $\xi = 1 / 10$ & $l = 64$ & $0.0183$ \\
%         \hline
%     \end{tabular}
%     \vspace{-2mm}
%     \caption{Different compression settings, where BPP is the average number of bits used per pixel.}
%     \label{tab:compression}
% \end{table}

\noindent\textbf{Results}. As shown in conditional video generation~\citep{Fangqiu25}, the work consistently outperforms baselines at all bitrates. Both metrics demonstrate improvements of 15–30\%, especially at extremely low bitrates, indicating that the generated videos better preserve perceptual quality, avoiding blocking artifacts (common in traditional codecs) and over-smoothed textures (typical of neural compression methods). Notably, under level 1 compression (with BPP below $0.007$), essential motion patterns and semantic content remain discernible, highlighting the potential of such a framework for bandwidth-constrained applications such as mobile streaming and surveillance.
% \autoref{fig:sota} summarizes the results across different compression levels. 

\subsection{Edge Video Analysis with Feature Encoding}\label{sec:2.2}
\subsubsection{Background}
With the rising need to run DNN-based applications on devices like wearables and Internet of Things (IoT) gadgets, edge AI has emerged as a critical approach for executing DNN tasks locally, closer to where data is generated~\citep{Guangxu20}. Among its various applications, edge video analytics~\citep{ananthanarayanan2017real_video_analytics} has seen significant growth and is now widely deployed across diverse domains~\citep{city20, othman2017new_surveillance, gao2018computer_healthcare}.

\noindent\textbf{Challenges}. To achieve the analytics process, edge systems typically transmit all input data from DNN-based models on edge devices to a nearby edge server for inference. When devices rely on wireless connections (\textit{e.g.}, autonomous vehicles and wearables), however, uplink bandwidth is constrained, and the quality of the communication channel can fluctuate significantly. Directly sending the large volumes of collected video data would result in excessive communication overhead.

To mitigate this, cooperative edge inference frameworks have been proposed~\citep{zhou2019edge_edge_intell,ShaoTWC}. In these frameworks, multiple edge devices capturing video data first extract compact input features that retain task-related information, and then transmit these features to an edge server for further processing. However, these approaches require high-complexity feature extraction, imposing heavy computational burdens on edge devices. \textbf{This inefficiency motivates approaches that trade local computation for reduced bandwidth, thereby lowering resource overhead while preserving task-relevant information}.

\noindent\textbf{Solutions}.
Related work develops a task-oriented framework to accelerate video data analytics at the network edge~\citep{ShaoZ20}. In particular, the work leverages the \textit{deterministic information bottleneck} principle~\citep{deterministic_ib} to extract features that are relevant to the task from each frame independently, while lowering the bitrate by utilizing previous features during the encoding process. This framework explicitly characterizes the trade-off between the relevance of the extracted features and the bandwidth consumption.

\subsubsection{Preliminary Results}
Existing work has shown the promise of feature encoding for video analysis on edge devices; for further details, see TOCOM-TEM~\citep{ShaoZ20}. Below, we briefly summarize the results.

\noindent\textbf{Tasks \& Datasets}. The study evaluates two tasks: 1)\textit{cross-camera pedestrian presence prediction} and 2) \textit{multi-view object detection}. It conducts the first task on Wildtrack~\citep{wildtrack_dataset}, and selects the EPFL MVMC Detection dataset~\citep{MVMC_dataset} for the latter one.

\noindent\textbf{Evaluation Metrics}.
The work focuses on the rate-performance trade-off that represents the upload overhead. For the cross-camera pedestrian presence prediction task, it reports the \textit{multiple object detection accuracy}~\citep{MODA_metric}; for multi-view object detection, it adapts the mean average precision metric to evaluate the accuracy of the predicted bounding boxes.

\noindent\textbf{Baselines}. The following data-oriented communication methods are utilized for comparison.
\begin{itemize}
\item\textit{Image compression}: Two conventional methods {JPEG} and {WebP}~\citep{si2016research_webp}, as well as two deep learning-based method (DIC)~\citep{balle2017end-to-end}.
% The two learning-based methods use neural networks to learn an encoder-decoder pair for image compression and decompression. The encoder outputs a quantized feature as a compact representation of the input image. An entropy model is learned end-to-end to reduce the bitrate of the quantized features via entropy coding.

\item\textit{Video compression}: AVC~\citep{richardson2011h_H264} and HEVC~\citep{sullivan2012overview_HECV}.
\end{itemize}

% \begin{figure}[t]
% \centering
% \begin{subfigure}[t]{0.48\linewidth}
%     \centering
%     \includegraphics[width=\linewidth]{figures/Final_wildtrack_MODA_bitrate_trade-off.pdf}
%     \caption{Rate-performance curves in the multi-camera pedestrian occupancy prediction task.}
%     \label{fig:Final_wildtrack_MODA_bitrate_trade-off}
% \end{subfigure}
% \hfill
% \begin{subfigure}[t]{0.48\linewidth}
%     \centering
%     \includegraphics[width=\linewidth]{figures/Final_EPFL_Accuracy_bitrate_trade-off.pdf}
%     \caption{Rate-performance curves in the multi-camera object detection task.}
%     \label{fig:Final_EPFL_Accuracy_bitrate_trade-off}
% \end{subfigure}
% \caption{Rate-performance curves of different methods for two tasks.}
% \label{fig:two_tasks_bitrate_trade-off}
% \end{figure}
\noindent\textbf{Results on Multi-Camera Pedestrian Occupancy Prediction}. Experimental results indicate that video compression methods consistently outperform image compression counterparts for data-oriented communication. This advantage arises from their ability to eliminate redundancy across consecutive frames, leading to higher compression ratios.

\noindent\textbf{Results on Multi-Camera Object Detection}. Data-oriented communication methods incur higher communication costs than task-oriented methods. This occurs because they ensure the accurate delivery of every transmitted bit without tailoring to the downstream task, resulting in a significant portion of the transmitted data being irrelevant for inference.

\subsection{Summary}
This section examines the computation-communication trade-off in modern AI systems, highlighting that increased computational effort can substantially reduce bandwidth consumption without compromising task performance or perceptual quality. As AI applications increasingly operate in edge–cloud environments, communication capacity has emerged as a critical bottleneck, often more restrictive than raw computational resources.

Through two representative scenarios, we demonstrate that shifting system design from data-centric transmission to computation-amplified reconstruction and inference. In video transmission (\S\ref{sec:2.1}), sparse yet semantically informative signals, combined with powerful generative models, enable high-quality video generation, particularly at extreme-low bitrates. In edge video analytics (\S\ref{sec:2.2}), task-aware feature extraction and temporal entropy modeling significantly reduce uplink communication by transmitting only information that is relevant to downstream inference tasks.

Together, these examples illustrate a unifying principle: \textbf{when bandwidth is scarce, computation can act as a powerful substitute for communication}. By exploiting learned priors, semantic representations, and task structure, AI systems can move beyond transmitting raw data and instead communicate only what is essential. This trade-off provides a flexible design paradigm for deploying intelligent services in bandwidth-constrained environments, enabling efficient system optimization.

\section{Trading Bandwidth for Memory}\label{sec:4}
\begin{figure}[h]
    % \vspace{-4mm}
    \includegraphics[width=\linewidth]{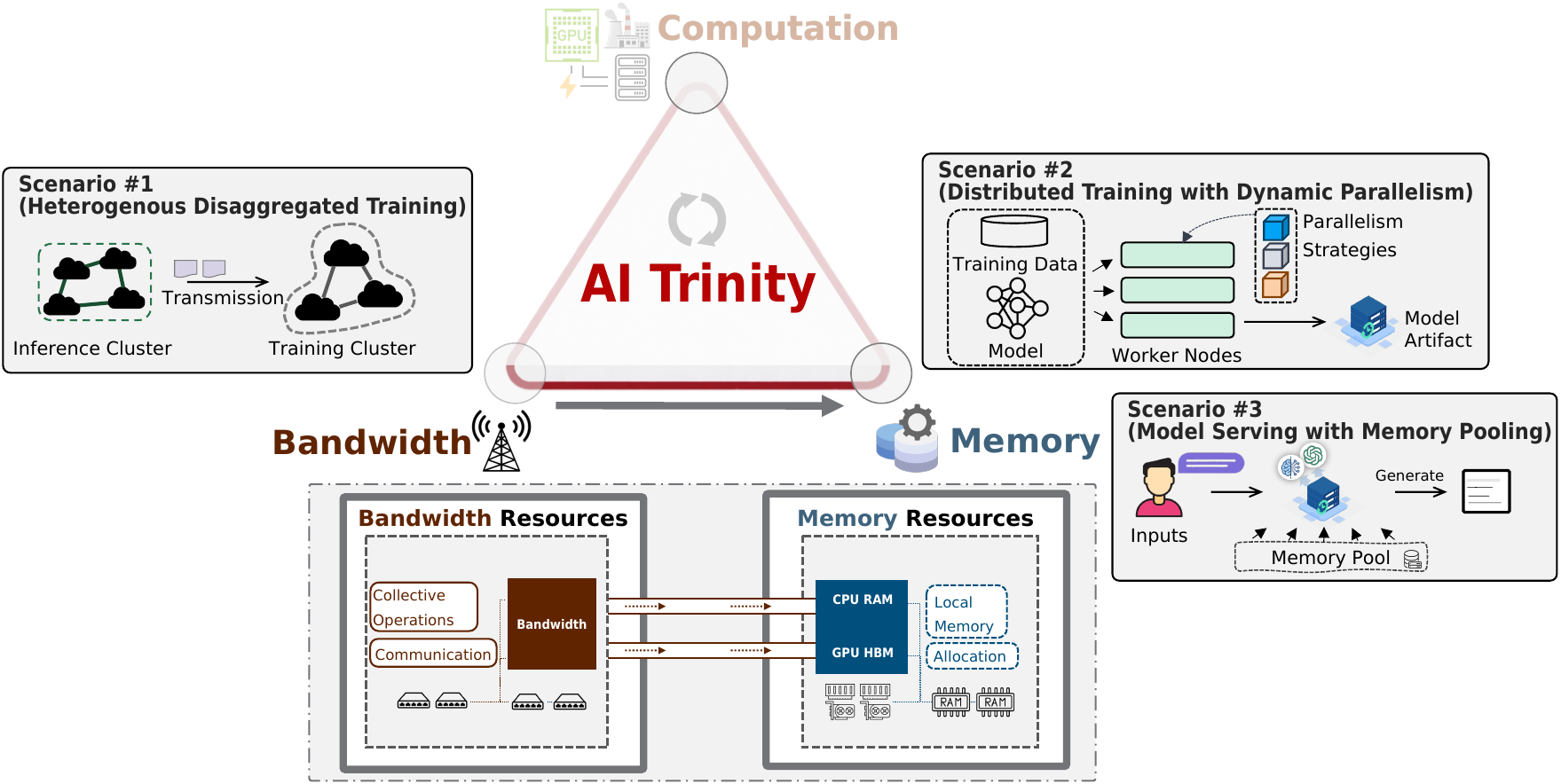}
    \caption{More \textbf{\textcolor{band}{Bandwidth}} for less \textbf{\textcolor{mem}{Memory}} trade-off.}
    \label{fig:band-mem}
\end{figure}

Nowadays, modern foundation models, including LLMs and their multimodal variants, are typically trained and served on accelerator clusters where on-device memory is limited relative to model size, batch size, and activation footprint. For example, storing the parameters of an $11$B foundation model alone requires roughly $40\text{ GB}$ of memory; when accounting for optimizer states, intermediate activations, and gradients, the total memory demand can easily exceed the capacity of a single accelerator. While memory-saving techniques such as activation checkpointing or parameter offloading can alleviate this pressure, they typically introduce additional computation or system complexity. 

An alternative approach is to leverage high-bandwidth interconnects to substitute memory locality with communication. A prominent example of this trade-off arises in data-parallel and model-parallel training. Techniques such as ZeRO-style optimizer~\citep{ZeRO20} state partitioning and fully sharded data parallelism distribute gradients, the states of optimizer, and model parameters across different devices. By sharding memory-intensive states, every device stores a fraction of the total model footprint, substantially reducing per-device memory requirements. The cost, however, is increased communication: parameters and gradients must be synchronized across devices at each iteration, often via all-reduce or all-gather operations. In this setting, abundant bandwidth and low-latency interconnects (\textit{e.g.}, NVLink or high-speed InfiniBand) are explicitly exploited to compensate for limited local memory. \textbf{This observation motivates approaches that deliberately trade bandwidth usage for reduced local memory consumption, enabling the development of large-scale models on memory-constrained devices without compromising system efficiency}.

\autoref{fig:band-mem} provides an overview of this trade-off by illustrating three representative scenarios. In the next subsections, we select two of these scenarios: we describe how memory consumption can be reduced through efficient collective operations under dynamic parallelism schemas in distributed training, and how heterogeneous training is employed to embody the principle of trading bandwidth for memory in large-scale model training.

\subsection{Efficient Generative Model Training with Dynamic Parallelism}\label{sec:3.1}
\subsubsection{Background}
 Recent years have witnessed a rapid evolution of AI-generated content across modalities including videos, images, and audio. As generative models transition from experimental creative tools to production-grade technologies, they are increasingly integrated into industrial pipelines, including short-form drama production, advertising, e-commerce, and digital entertainment. This widespread adoption underscores an unprecedented surge in demand for efficient generative models.

 Modern generative models (\textit{e.g.}, video generation models) must simultaneously handle long temporal sequences, high spatial resolutions, and complex multimodal inputs, leading to extreme computational and memory pressure during training. In practice, training state-of-the-art generative models often requires thousands of GPUs running continuously for weeks. Without effective system-level optimizations, such training pipelines incur prohibitive costs, exhibit low hardware utilization, and are inaccessible to small and medium-sized enterprises and independent developers.

 \noindent\textbf{Challenges}. Distributed training is therefore fundamental to the feasibility of generative model development, as it leverages inter-device bandwidth to aggregate memory capacity beyond that of a single device. While existing distributed frameworks (\textit{e.g.}, PyTorch FSDP~\citep{FSDP23} and DeepSpeed~\citep{ZeRO20}) have achieved success in training LLMs, they exhibit significant limitations when applied to generative models. For example, existing frameworks struggle to support long video sequences beyond a few seconds at $720$P resolution, frequently encountering GPU memory exhaustion even when additional computational resources are available. \textbf{This growing demand motivates approaches that trade inter-device bandwidth for single-device memory, thus enabling more efficient training systems, particularly for modern generative models}.

 \noindent\textbf{Proposed Method}. We propose a novel solution designed for efficient large-scale generative model training. The key innovation is to decompose the training process into fundamental computational units and apply dynamic parallelism configurations for fine-grained orchestration. Additionally, we perform topology-aware mapping of these units onto specific devices. This approach leverages high-bandwidth interconnects to offload memory requirements across multiple devices, minimizing unnecessary communication overhead and hence maximizing overall training efficiency.
% The trade-off between bandwidth usage and memory consumption has become a critical consideration in the development of large-scale distributed training systems. As the size of modern generative models increases, traditional training techniques that rely on local memory alone are reaching their limits. To address this, more sophisticated distributed training approaches have emerged that 

\subsubsection{Exploring the Potential of the Trade-offs}
This section explores the theoretical implications of memory-bandwidth trade-off, focusing on how much communication can compensate for limited memory in distributed training scenarios.

Let the total memory required to train a model be $M_{\text{total}}$. Each device has a memory capacity of $M_{\text{avail}}$, and the training is distributed across $N$ devices. When the aggregate device memory is insufficient, a \textit{memory deficit} arises:

$$
M_{\text{miss}} = \max\left(0, M_{\text{total}} - N \cdot M_{\text{avail}}\right).
$$
where $M_{\text{miss}}$ denotes the deficit that represents the portion of model states that cannot be stored locally and must instead be exchanged through communication during training.

In data-parallel training, collective operations (\textit{e.g.}, \textit{all-reduce}) are used to synchronize parameters, gradients, and optimizer states. The total communication cost per iteration consists of the standard synchronization cost plus an additional cost proportional to the memory deficit:

$$
C_{\text{comm}} = T_{\text{all-reduce}} \cdot (P + G + O) + k \cdot M_{\text{miss}},
$$
where $P$, $G$, and $O$ denote the gradients, optimizer states, and also model parameters, respectively, and $k$ is a system-dependent factor representing the additional bandwidth required to compensate for $1$ unit of missing memory.

\noindent\textbf{An Illustrative Example}. Assume the total required memory of training a model is $M_{\text{total}} = 400\text{ GB}$, and each device has $M_{\text{avail}} = 100\text{ GB}$. With $N = 2$ devices, the memory deficit is $M_{\text{miss}} = 400 - 2 \cdot 100 = 200\text{ GB}$. If the system requires $k = 0.5$ units of bandwidth per unit of missing memory, the extra communication cost roughly becomes $0.5 \cdot 200 = 100\text{ GB per iteration}$. This shows that $100\text{ GB}$ of bandwidth per iteration can effectively replace $200\text{ GB}$ of missing memory.

\subsubsection{Preliminary Results}\label{sec:3.1.3}
We conduct preliminary experiments to train the text-to-video (T2V) model HunyuanVideo, a $13$B-parameter model~\citep{HunyuanVideo24}. Experiments are carried out on a two-node GPU cluster interconnected by a $400\text{ Gbps}$ InfiniBand network.

\noindent\textbf{Experimental Workloads}. We utilize publicly available datasets and adopt a \textit{three-stage} training strategy, with each stage gradually increasing the data scale and computational complexity:
\begin{itemize}
    \item\textit{Stage $1$} uses the WebVid dataset~\citep{WebVid21} with about $7$ million clips at a $360$P resolution.
    \item\textit{Stage $2$} uses Koala~\citep{Koala25} with about $20$ million clips at a higher $720$P resolution.
    \item\textit{Stage $3$} scales up an internal dataset (named Lynx) that contains approximately $10$ million clips at $1080$P to represent a production-level workload.
\end{itemize}

\noindent\textbf{Baselines}. We use two state-of-the-art distributed training systems, namely DeepSpeed~\citep{ZeRO20} and Megatron-LM~\citep{MegatronLM19}. Both baselines employ a traditional static parallelism paradigm and leverage data-centric parallelisms.

\begin{figure}[h]
    \centering
    % \vspace{2mm}
    \includegraphics[width=0.95\textwidth]{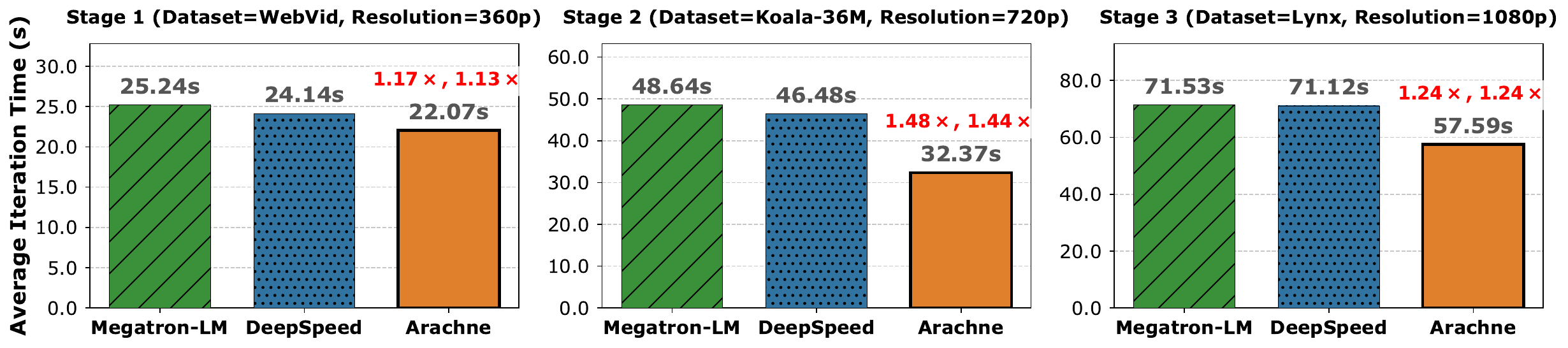}
    % \vspace{-2mm}
    \caption{Average iteration time (in seconds) on the three different training workloads.}
\label{fig:avg_iteartion_time_across_stage}
\end{figure}

\noindent\textbf{Results}. The end-to-end training results are presented in~\autoref{fig:avg_iteartion_time_across_stage}. Across all three stages, our proposed framework (named Arachne) consistently outperforms both baselines, with its performance advantage becoming more pronounced from Stage $1$ to Stage $2$. Since data heterogeneity and computational demands intensify, the resulting workload imbalance amplifies the benefits of the fine-grained execution, yielding peak speedups of 48\% over Megatron-LM and 44\% over DeepSpeed.

% It's worth noting that the performance advantage moderates in Stage 3, though Arachne still achieves a 24\% speedup. This moderation actually arises from hardware memory constraints that cap the maximum sequence length at only $57$ frames. Thus, training is restricted to a relatively homogeneous $6.5\%$ fraction of the dataset. The reduced heterogeneity substantially alleviates load imbalance, thereby narrowing the scope for further optimization by Arachne. This effect produces the ``V-shaped'' performance trend observed across the entire training process.

\subsection{Heterogeneous Training with Messaging}\label{sec:3.2}
\subsubsection{Background}
  \begin{figure}[h]
    \centering
    % \vspace{-2mm}
    \includegraphics[width=.7\linewidth]{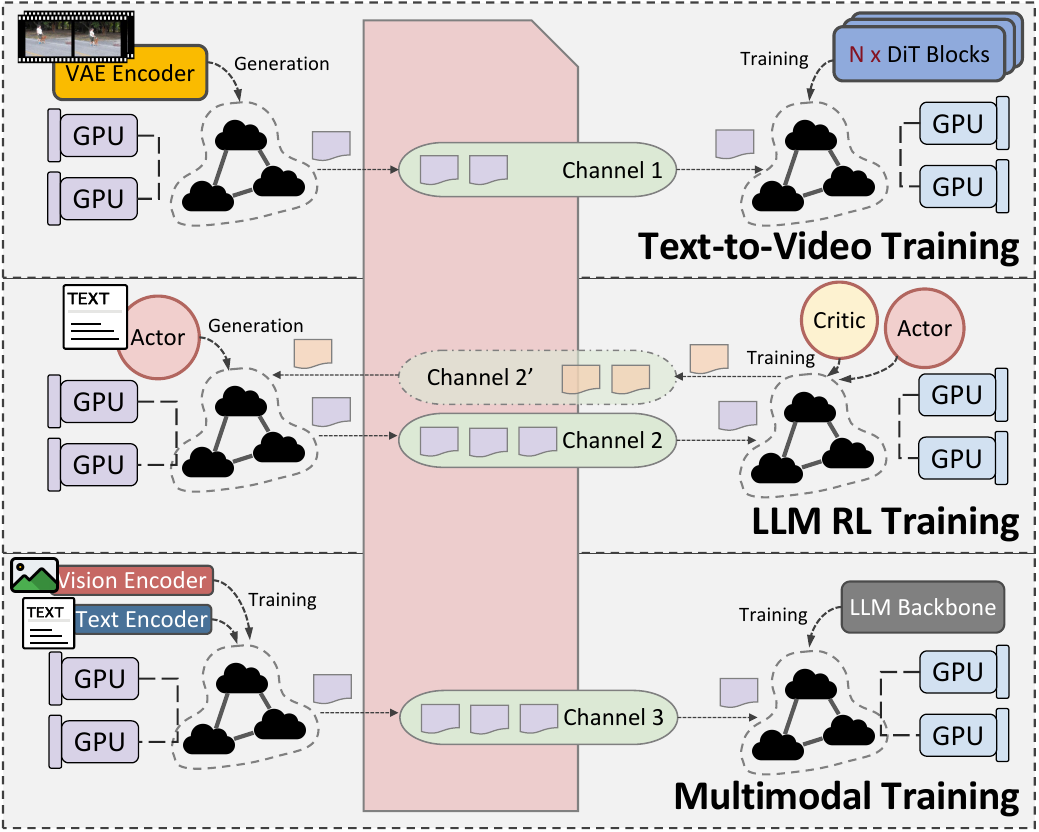}
    % \vspace{-2mm}
    \caption{Three representative scenarios of heterogeneous disaggregated training across clusters.}
    \label{fig:disaggregated-training}
\end{figure}

 Training large-scale models necessitates the deployment of clusters with thousands of GPUs~\citep{gpt20, Seedance25}. To facilitate such training, most existing systems are naturally built on the colocated paradigm, in which different training components are mapped onto the same set of GPUs. Unfortunately, these systems suffer from resource coupling (\textit{i.e.}, a tight interdependence among GPUs, model components, and training stages) as training scales, leading to significant resource underutilization~\citep{StreamRL25}. In contrast, \textit{heterogeneous training paradigm}, which decouples training components from specific hardware across clusters to utilize the most suitable resources, is increasingly acknowledged. \autoref{fig:disaggregated-training} illustrates three representative training cases across T2V model ~\citep{stepvideo25}, LLM reinforcement learning (RL)~\citep{AsyncFlow25}, and multimodal LLM~\citep{DistTrain24}. Indeed, shifting to a heterogeneous disaggregated architecture enables dynamic resource provisioning, supports diverse hardware setups, and enables cross-cluster deployment.

\noindent\textbf{Challenges}. To support this paradigm, most existing implementations adopt relatively simple designs that focus primarily on the communication layer, either by leveraging existing communication protocols~\citep{StreamRL25} or by developing communication frameworks~\citep{stepvideo25}. While a recent effort~\citep{AsyncFlow25} extends this direction by incorporating a distributed storage layer to provide unified data management, its design is largely limited to RL scenarios. \textbf{We contend that fully harnessing the potential of the heterogeneous training paradigm requires a unified system that combines both communication and memory optimizations, exploiting the communication-memory trade-off to maximize overall system performance}.

\noindent\textbf{Proposed Method}. We present a novel framework tailored for the heterogeneous training paradigm. Its core idea is to leverage a message queue system for efficient data transmission across clusters. By exploiting inter-cluster bandwidth, our framework alleviates single-machine memory constraints in model training and enables the effective use of lower-end computational resources to complement high-end hardware, thereby improving overall resource efficiency and reducing training cost.

% In this section, we conduct preliminary experiments to explore the potential system inefficiencies under heterogeneous training and envision the development of a more efficient system that embodies the principle of trading bandwidth for reduced memory usage.

\subsubsection{Preliminary Results}
We conduct preliminary experiments on the T2V training scenario using the video generation model WAN2.2~\citep{wan25}, carried out on two separate clusters: a low-end GPU cluster serving as ``\textit{producer}'' for VAE forward passes, and a high-end cluster acting as ``\textit{consumer}'' for DiT training.

\noindent\textbf{Data}. For simplicity, we simulate diverse video data at frame rates between $5$ fps and $129$ fps, all at a $720$P resolution, for the purpose of evaluation.

\begin{figure}[h]
    \centering
    \includegraphics[width=.8\linewidth]{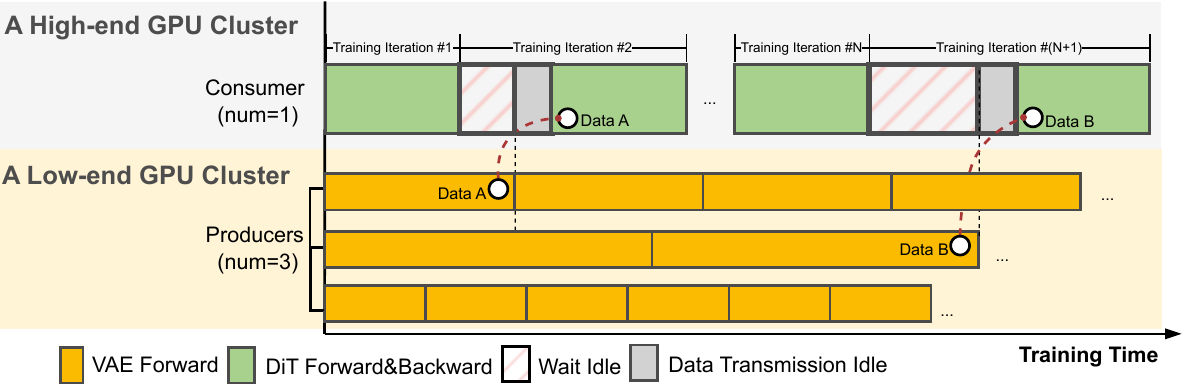}
    % \vspace{-2mm}
    \caption{Training steps of the T2V model with two separate clusters.}
    \label{fig:heterogeneous}
\end{figure}

\noindent\textbf{Results}. By enabling heterogeneous training across two clusters, our framework can substantially reduce computational demands: four low-end GPUs can replace a single high-end GPU, compared with a colocated training setup on the high-end cluster. In addition, the framework supports a multi-process data messaging mechanism that breaks the single-thread bandwidth limit of $50$ MB/s in our cluster setup. With a $10\text{ Gbps}$ network, our proposed framework can support heterogeneous training across approximately $20$ nodes.

\autoref{fig:heterogeneous} provides a snapshot of the training procedure, illustrating the dynamic interaction between high- and low-end nodes. We observe that system ``bubbles'' may occur when the dynamics of the data distribution. Future optimizations, such as dynamic parallelism conditioned on fixed bandwidth restriction, could further improve scalability across even larger clusters.

\subsection{Summary}
This section examines the trade-off with bandwidth and memory consumption, particularly in the context of training modern large-scale generative models. We described two scenarios in model training: first, a framework that dynamically orchestrates computational tasks and uses high-bandwidth interconnects to optimize memory usage (\S\ref{sec:3.1}), and second, the benefits of heterogeneous training for efficiently scaling large models across diverse hardware (\S\ref{sec:3.2}). These approaches demonstrate how strategic use of bandwidth-memory trade-offs improves the efficiency of model training.

In a nutshell, the use cases, particularly in training scenarios, highlight the basic system design principle: \textbf{bandwidth can be harnessed to coherently aggregate memory resources into a unified memory fabric, thereby overcoming the limitations of local storage capacity}. As generative models continue to scale, it is essential to develop systems that effectively balance memory usage and communication bandwidth. The techniques proposed in this section provide a promising direction for addressing these challenges, enabling more efficient training of modern large-scale AI models.

\section{Trading Memory for Computation}\label{sec:5}
A key driver towards better AI models has been modern computing. From their pen-and-paper origins, AI models have transformed in capacity and predictive power by the exponential rise in computing~\citep{Moore65}. But the computational cost of developing large-scale AI models is substantial and increases with model size. Training these frontier models often requires thousands of accelerator-hours on specialized hardware, resulting in high financial cost and energy consumption~\citep{Gopher21, LaMDA22, Chinchilla22}. Moreover, inference costs can rival or exceed training costs in real-world deployments. This challenge is particularly pronounced with the emergence of diffusion-based generative models (\textit{e.g.}, diffusion transformers (DiT))~\citep{openai2024sora, wan25, Seedance25}, whose iterative denoising procedures require multiple forward passes and therefore incur substantial computational overhead.

\begin{figure}[h]
    \vspace{-2mm}
    \includegraphics[width=\linewidth]{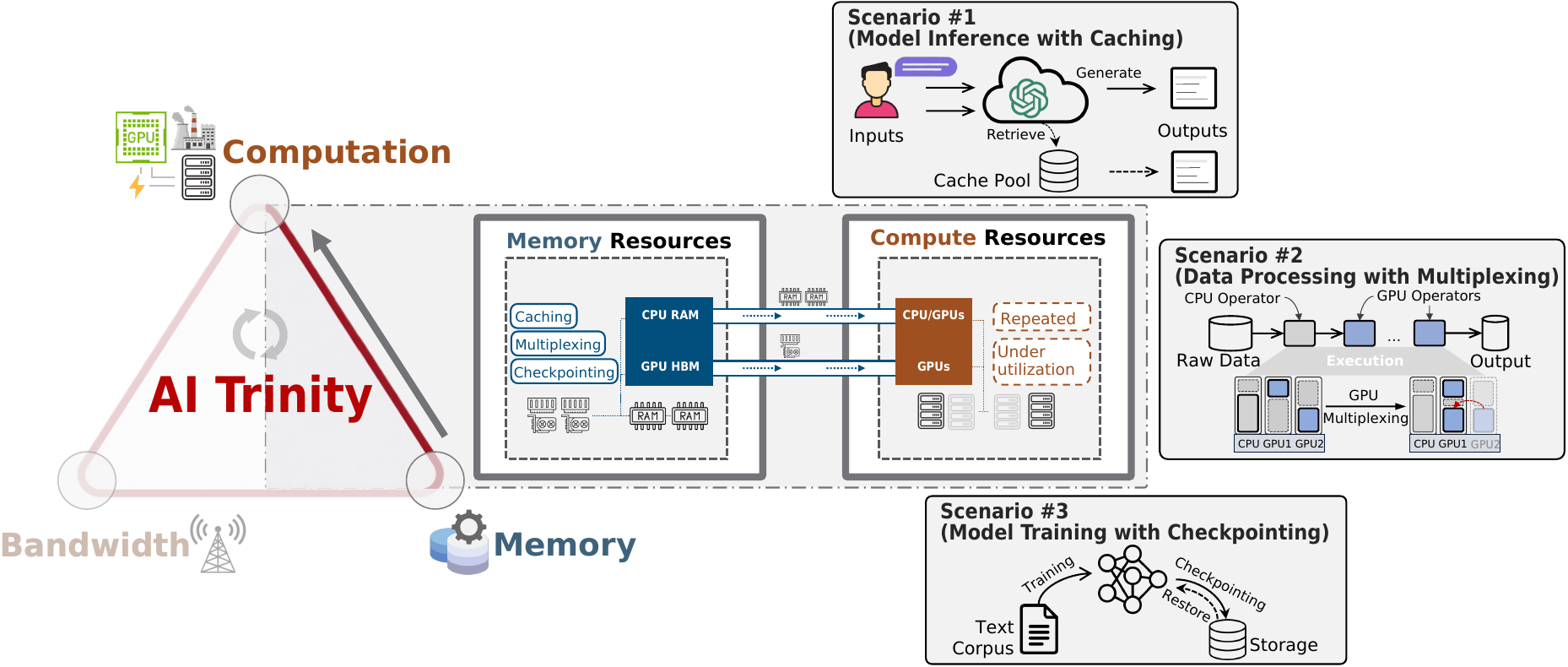}
    \vspace{-7mm}
    \caption{More \textbf{\textcolor{mem}{Memory}} for less \textbf{\textcolor{comp}{Computation}} trade-off.}
    \label{fig:mem-comp}
\end{figure}

We report the FLOPs required for these models as illustrative examples. Training an LLM with $6.8$B parameters using the common approximation\footnote{$C$ denotes floating point operations (FLOPs), $D$ the number of training tokens, and $N$ the number of model parameters.} $C = 6DN$~\citep{Scaling20} requires approximately $5.76\times 10^8$ petaFLOPs~\citep{Chinchilla22}; serving an inference request with a $70$B-parameters LLM~\citep{Llama3-24}, using the approximation\footnote{Following~\citep{Scaling20}, we assume that the backward pass incurs twice the FLOPs of the forward pass; inference therefore accounts only for the forward pass.} $C = 2DN$, and processing ten thousand tokens, requires roughly $1.4$ petaFLOPs, whereas generating a $5$-second video at $16$ fps with a resolution of $1056 \times 1932$ using a 14B-parameter DiT model~\citep{wan25} requires up to about $30$ petaFLOPs.

% Assuming an NVIDIA H100 SXM GPU with FP16 Tensor Core peak throughput of $1$ petaFLOPs (\textit{i.e.}, $1$ petaFLOPs per second) at 100\% utilization, completing such an inference request would take approximately $1.4$ seconds. Hyperbolic Labs cost: \$1.49/hr with no additional fees or operational overhead, costs roughly $0.0000662\$$

% For instance~\citep{Mooncake25}, the KV cache size for Llama 3.1 8B [21] is approximately 1.2 GB for a single request with ten thousand tokens.

However, in contrast to the rapidly increasing cost of computation, the cost of storage has decreased exponentially since the 1950s. In addition, improvements in storage capacity and computational power have progressed at markedly different rates. Kryder's law~\citep{KrydersLaw05} observes that storage density has historically doubled roughly every year, while Moore's law~\citep{Moore65} indicates that the transistor density achievable on a chip within reasonable cost doubles around every two years. \textbf{This growing asymmetry motivates approaches that embody the principle of trading more storage in place of less computation, thereby enabling more efficient AI systems under storage-rich but compute-constrained scenarios}.

\autoref{fig:mem-comp} provides an overview of this trade-off by illustrating three representative scenarios. In the following two subsections, we focus on two of these scenarios in detail, explaining how computation can be effectively reduced through caching without compromising quality in real-time video generation, and how memory sharing can reduce computational overhead in modern, sophisticated data-processing pipelines.

\subsection{Accelerate Video Diffusion Inference with Caching}\label{sec:4.1}
\subsubsection{Background}
In recent years, diffusion-based generative models~\citep{song2019generative, ho2020denoising, dhariwal2021diffusion} have seen notable success in visual generation tasks, emerging as a cornerstone paradigm in AI-driven creative systems. They enable a broad range of applications, from digital content creation and interactive media to cinematic production. Building upon these advances, DiT-based~\citep{DiT23} T2V models have progressed rapidly, with representative models such as Sora~\citep{openai2024sora}, Wan~\citep{wan25}, and SeedDance~\citep{Seedance25}. In the T2V generation setting, a diffusion model takes a text prompt describing the desired content, initializes the video representation with pure Gaussian noise, and then iteratively predicts and removes noise through a series of denoising steps, synthesizing the final video via a Markov process~\citep{Markov17}.

\noindent\textbf{Challenges}. However, this inference process suffers from substantial computational complexity. Traditional diffusion models require up to $1000$ iterative denoising steps~\citep{GAN21} for this; even with modern optimizations (\textit{e.g.}, feature caching discussed below), mainstream pretrained models~\citep{wan25, Seedance25} reduce this to only about $50$ iterative steps\footnote{Although distillation methods~\citep{T2V-Turbo24, DOLLAR24} can achieve few-step video generation, they rely on costly post-training procedures and are therefore excluded from our consideration.}, which still results in several minutes to generate a single video. For example, generating a $2$-second $64$-frame video at a resolution of $320\times 512$ requires about $240$ seconds using a single A100 GPU, whereas producing a $20$-second $24$-fps 2K video requires more than $70$ minutes using an H100 GPU.

A promising direction for improving inference efficiency is \textit{caching}, which begins the generation process using partially computed results instead of starting from pure Gaussian noise, thereby reducing redundant computation. Existing works have proposed various caching techniques to accelerate video diffusion inference~\citep{Cache-Me24, Learning-to-Cache24, DeepCache24, TaylorSeers25, flexcache25, MoDM25, OmniCache25}, largely focusing on feature caching, which reuses DiT features from earlier timesteps to skip redundant computations in later timesteps within an inference request. However, as noted above, the potential of these approaches is inherently limited.

In contrast, few works~\citep{NIRVANA24, flexcache25} explore \textit{approximate caching} strategies\textemdash analogous to context caching in LLMs~\citep{SGLang23, MemServe24}\textemdash that reuse early denoising latents across similar inference requests to shorten or even skip the sampling process under limited memory budgets. Despite their promise, effective solutions remain largely underexplored in the T2V setting\footnote{We attempted to reproduce the results reported from~\citet{flexcache25} but failed due to missing implementation details.}. Thus, \textbf{these limitations motivate approaches that leverage modern caching techniques to more fully explore the memory-computation trade-off in video diffusion inference}.

\noindent\textbf{\twemoji[height=1.0em]{thinking face}Scaling Laws in the Memory-Computation trade-off}? In this section, we investigate the potential of replacing computation with storage by expanding caching capacity, with a particular focus on how \textit{scaling} caching memory affects the system efficiency of video diffusion inference. Drawing on findings from text-to-image (T2I) generation~\citep{NIRVANA24}, we characterize the potential scaling behavior in the T2V setting and empirically examine the practical challenges that hinder effective scaling. We hope this analysis provides actionable insights for leveraging caching effectively to reduce inference time in real-world video diffusion applications.

\subsubsection{Exploring the Potential of the Trade-offs}
This section presents a theoretical analysis of how additional memory (cache capacity) can compensate for computational requirements during video diffusion inference.

Let a diffusion-based T2V model perform $S$ denoising timesteps during a full generation.  Denote the per-step compute cost (FLOPs) by $c_{\mathrm{step}}$, so the cost of a full, uncached generation is
$$
C_{\mathrm{full}} = S \cdot c_{\mathrm{step}}.
$$

A cache entry stores a saved {\em latent} obtained after $r$ denoising steps (we call $r$ the reuse depth): if a request hits that entry, the generator can begin from that latent and only execute the remaining $S-r$ steps.  Let the (average) size of a cache entry be $s_e$, and let the total cache capacity be $M$, so the (max) number of entries storable is roughly $N(M)=\lfloor M/s_e\rfloor$.

We model an online workload where similar inference requests recur. The cache hit probability under a given caching policy $\pi$ (which specifies the similarity threshold, replacement policy, compression level, \textit{etc}) is denoted by $h(M; r, \pi)$.
Finally, using a cached latent from depth $r$ saves,
$$
c_{\mathrm{saved}}(r) = r \cdot c_{\mathrm{step}}.
$$

With this notation, the expected compute cost per request when using the cache is
\begin{equation}
\label{eq:cost-simplified}
\mathbb{E}[C](M;r,\pi)
= C_{\mathrm{full}} - h(M;r,\pi) \cdot c_{\mathrm{saved}}(r).
\end{equation}

\noindent\textbf{Trade-offs}. All concrete trade-offs pass through the hit-rate function $h(M;r,\pi)$. Depending on the workload distribution and the caching policy, two simple families capture common empirical behavior,

\begin{itemize}
\item\textbf{Exponential Saturation} (fast initial returns followed by a plateau):
\[
h(M) = 1 - e^{-\beta(r,\pi) M / s_e}.
\]
where $h(M)$ is an empirical measurement on a representative log. Substituting into \eqref{eq:cost-simplified} yields,
\[
\mathbb{E}[C](M)
= C_{\mathrm{full}} - \big(1 - e^{-\beta M/s_e}\big) \cdot c_{\mathrm{saved}}(r).
\]
The marginal time benefit per additional byte of cache decays exponentially:
\[
\frac{d\mathbb{E}[C]}{dM}
= -\frac{\beta}{s_e} e^{-\beta M/s_e}\, \cdot c_{\mathrm{saved}}(r).
\]
This directly shows diminishing returns: after a few characteristic bytes $\sim s_e/\beta$, additional memory yields negligible latency improvements.

\item\textbf{Power-Law} (long tail of rare hits):
$$
h(M) = 1 - (1+\kappa, M)^{-\gamma},\quad \gamma>0.
$$
Here, the marginal improvement falls off polynomially, and large increases in $M$ can still produce nontrivial gains if the workload has a heavy tail of repeated patterns.
\end{itemize}

\noindent\textbf{An Illustrative Example}. Consider a model requiring $S=50$ denoising steps per video, with per-step cost $c_{\mathrm{step}} \approx 10^9$ FLOPs. Suppose each cache stores a latent after $r=20$ steps, with size $s_e = 2\text{ GB}$. If the total cache capacity is $M=10\text{ GB}$, roughly $N(M) = 5$ entries can be stored. Assuming the caching policy achieves a hit rate $h(M)=0.6$, the expected compute saved per request is:
$$
h(M) \cdot c_{\mathrm{saved}}(r) = 0.6 \times 20 \times 10^9 = 1.2 \times 10^{10}\ \text{FLOPs}.
$$
This shows that allocating $10\text{ GB}$ of cache allows the system to save $1.2\times 10^{10}$ FLOPs per inference on average. In other words, in this setting, every $\sim 1.7\text{ GB}$ of memory yields $\sim 2\times 10^9$ FLOPs of saved computation, illustrating a direct trade-off between memory capacity and compute requirements.

\subsubsection{Preliminary Results}
We conduct experiments to study whether scaling storage capacity can lead to greater computational savings in the diffusion-based model inference scenario.

\begin{wraptable}{r}{0.35\textwidth}
    \vspace{-10mm}
    \centering
    \small
    \begin{tabular}{cc}
        \hline
        Similarity Score $s$ & Reuse Depth \\
        \hline
        $s > 0.95$          & 25 \\
        $0.9 < s \le 0.95$  & 20 \\
        $0.85 < s \le 0.9$  & 15 \\
        $0.75 < s \le 0.85$ & 10 \\
        $0.65 < s \le 0.75$ & 5  \\
        $s \le 0.65$        & 0  \\
        \hline
    \end{tabular}
    \vspace{-3mm}
    \caption{Cache retrieval settings.}
    \label{tab:cache_retrieval}
\end{wraptable}
\noindent\textbf{Dataset \& Model}. We select the publicly available VidProM dataset~\citep{VidProM24}, which contains text prompts from real-world Discord users, each associated with the timestamp of its request. We employ Wan2.2~\citep{wan25} as the underlying inference serving model to generate videos.

\noindent\textbf{Experimental Settings}. We adapt NIRVANA~\citep{NIRVANA24} to perform video caching\footnote{The evaluation is conducted based on the assumption that the strategies and settings used for T2I generation can also be applied to video generation scenarios.}. The following are the detailed settings for this evaluation:

\begin{itemize}
    \item \textit{Cache Saturation \& Replacement}. We simulate requests based on their associated timestamps, using early requests to populate the cache under varying storage capacities. Once it reaches the capacity, we apply the standard least recently used eviction policy to maintain cache efficiency.
    
    \item\textit{Cache Selection}. We store intermediate-states (\textit{i.e.}, latents) at five distinct denoising steps (\textit{i.e.}, $5^{th}$, $10^{th}$, $15^{th}$, $20^{th}$, $25^{th}$). The size of each latent is $0.016$ GB, $0.04$ GB, and $0.07\text{ GB}$ for different targeted resolutions ($720$P, $1080$P, and $2$K), respectively.
    
    \item\textit{Cache Retrieval}. For each prompt, we convert it to a CLIP embedding~\citep{CLIP21} (with $768$ dimensions) and compute its similarity score. Based on the similarity score, we determine the reuse depth using the settings in~\autoref{tab:cache_retrieval}.
\end{itemize}

\definecolor{graybox}{RGB}{90,90,90}
\begin{wrapfigure}{r}{0.5\textwidth}
    \vspace{-25pt}
    \begin{center}
    \includegraphics[width=\linewidth]{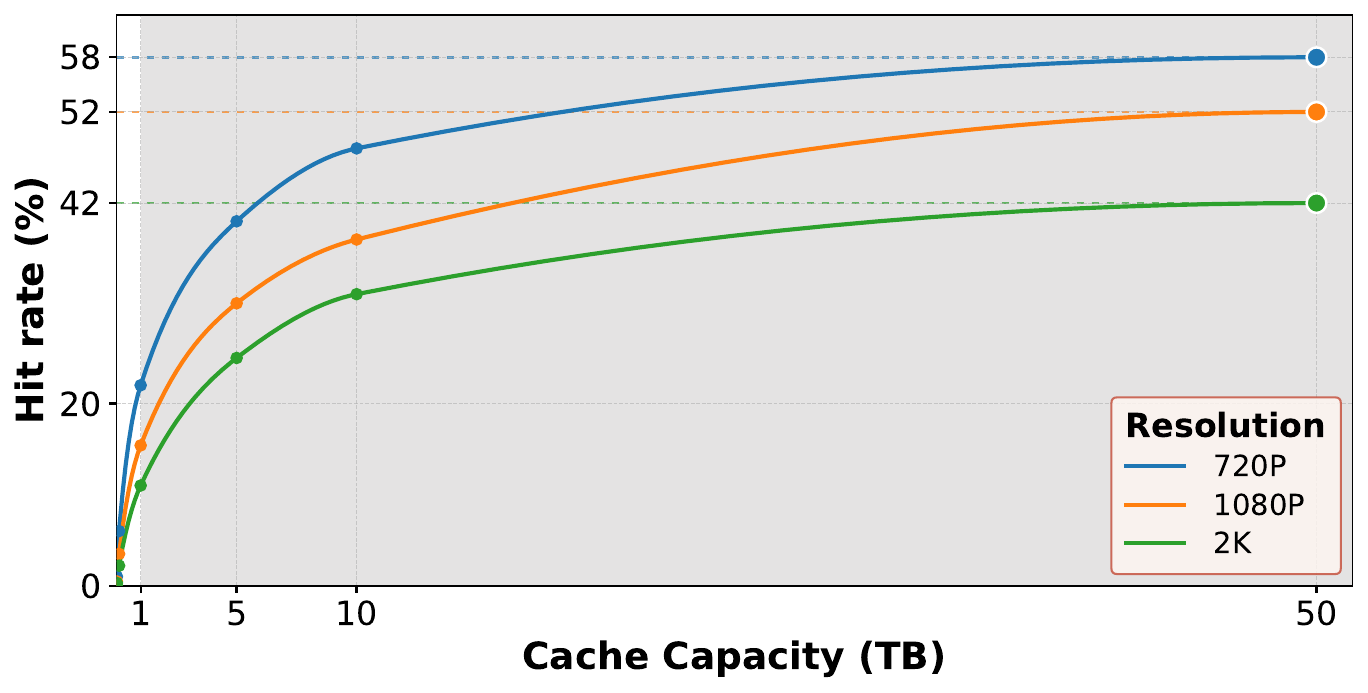}
    \end{center}
    \vspace{-1.8em}
    \caption{Scaling behavior of hit rate versus capacity at different resolutions. \textcolor{graybox}{\textbf{Gray region}} highlights significant unexplored scaling potential beyond current studies (with a $1$TB maximum).} 
    \label{fig:scaling}
\end{wrapfigure}
% \begin{figure}[h]
%     \centering
%     \vspace{-3mm}
%     \includegraphics[width=0.7\linewidth]{figures/scaling.pdf}
%     \vspace{-2mm}
%     \caption{Scaling behavior of hit rate versus capacity at different resolutions. \textcolor{graybox}{\textbf{Gray region}} highlights significant unexplored scaling potential beyond current studies (with a $1$ TB maximum).} 
%     \label{fig:scaling}
% \end{figure}

\noindent\textbf{Results}. \autoref{fig:scaling} illustrates the potential scaling behavior in the T2V setting, targeting different generation resolutions. The results show a sublinear scale trend as the cache capacity is gradually increased. In particular, when the capacity reaches $50$ TB, the serving system achieves a 58\% hit rate for generating $720$P videos, leading to significant computational savings for video diffusion inference. On the other hand, despite a relatively lower hit rate (42\%) for high-quality 2K videos due to larger latent sizes (\textit{i.e.}, fewer caches being maintained), further scaling the memory capacity should yield greater benefits in such scenarios. 

\begin{figure}[h]
\centering
\vspace{-20pt}
\resizebox{\linewidth}{!}{
\begin{tabular}{p{9.3cm}|p{3.1cm}}
& {\textbf{\scriptsize Videos: click to play in Adobe Acrobat}}\\[-2pt]
\vbox{
\hbox{\zgfrowlabel{w/o cache}%
\includegraphics[width=0.333\linewidth]{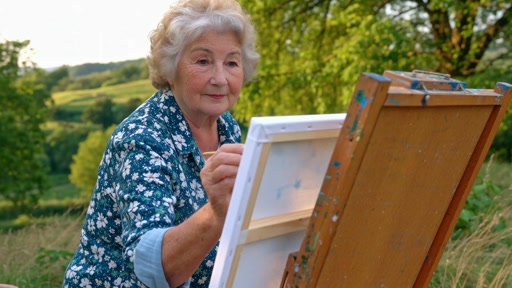}%
\includegraphics[width=0.333\linewidth]{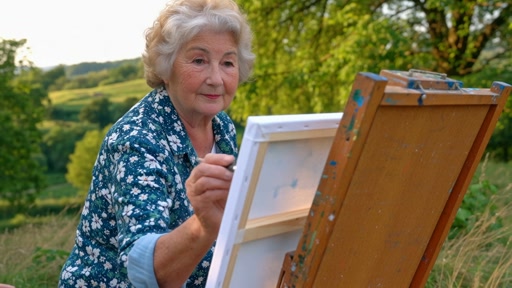}%
\includegraphics[width=0.333\linewidth]{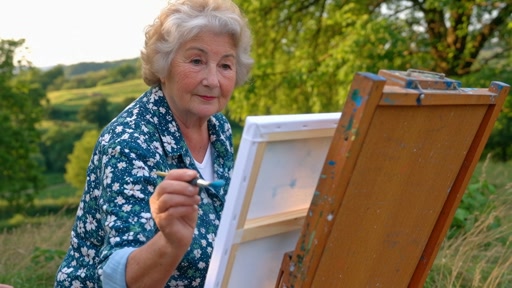}}%
\hbox{\zgfrowlabel{w/ cache}%
\includegraphics[width=0.333\linewidth]{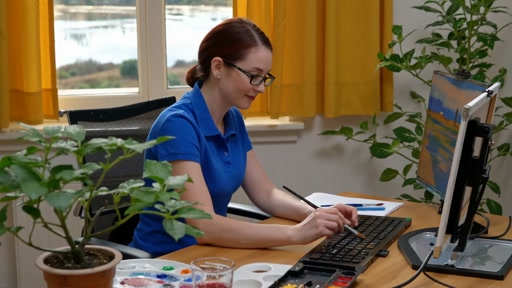}%
\includegraphics[width=0.333\linewidth]{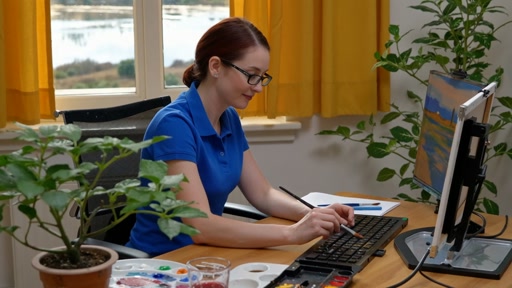}%
\includegraphics[width=0.333\linewidth]{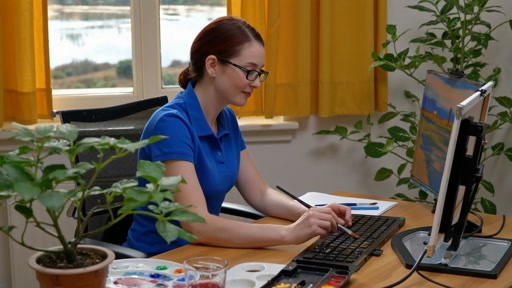}}%
\hbox{\zgfrowlabel{ref}%
\includegraphics[width=0.333\linewidth]{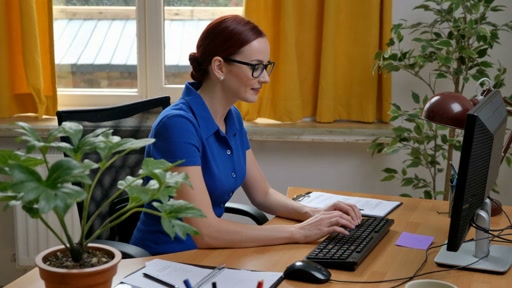}%
\includegraphics[width=0.333\linewidth]{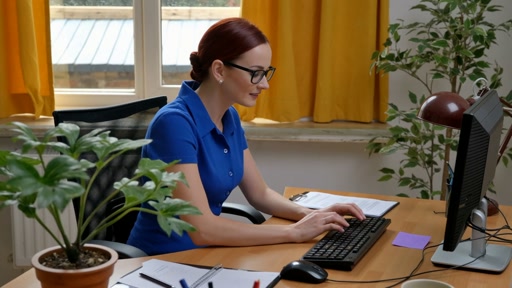}%
\includegraphics[width=0.333\linewidth]{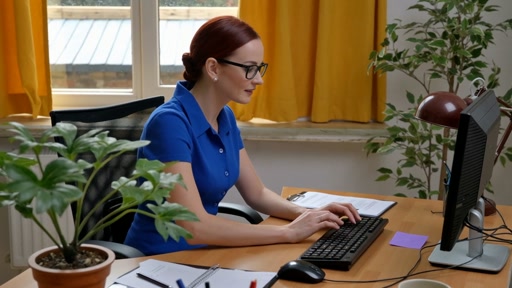}}%
\vspace{5pt}
\hbox{\zgfrowlabel{w/o cache}%
\includegraphics[width=0.333\linewidth]{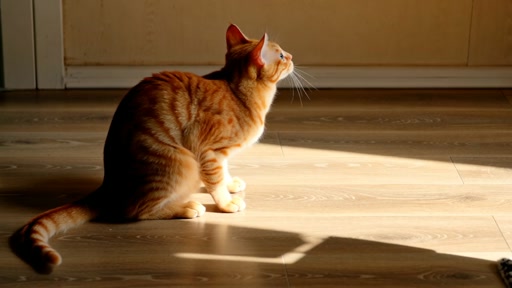}%
\includegraphics[width=0.333\linewidth]{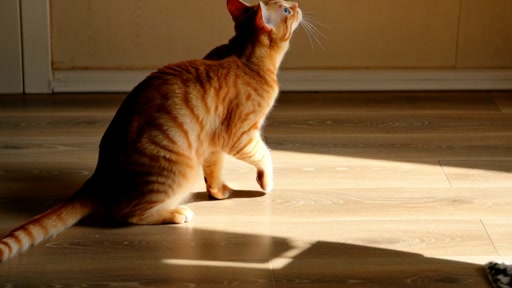}%
\includegraphics[width=0.333\linewidth]{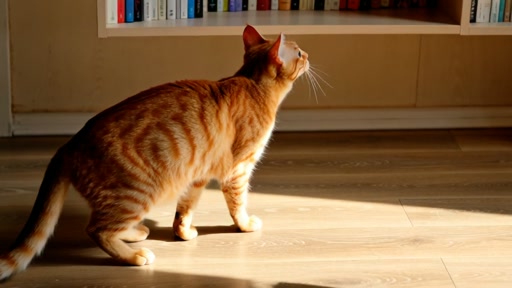}}%
\hbox{\zgfrowlabel{w/ cache}%
\includegraphics[width=0.333\linewidth]{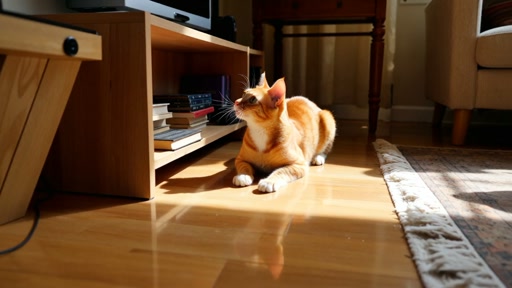}%
\includegraphics[width=0.333\linewidth]{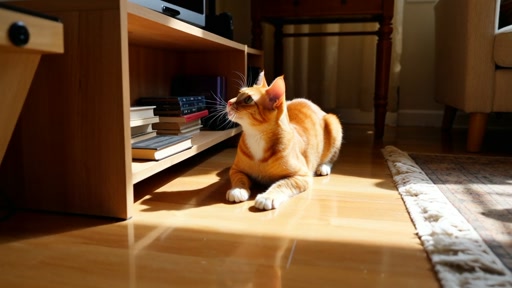}%
\includegraphics[width=0.333\linewidth]{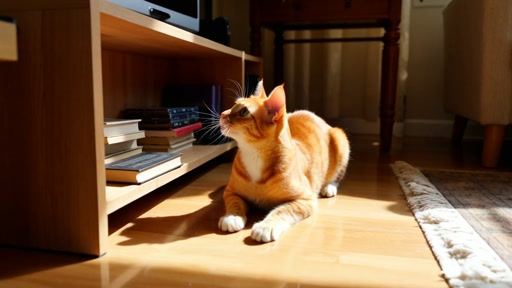}}%
\hbox{\zgfrowlabel{ref}%
\includegraphics[width=0.333\linewidth]{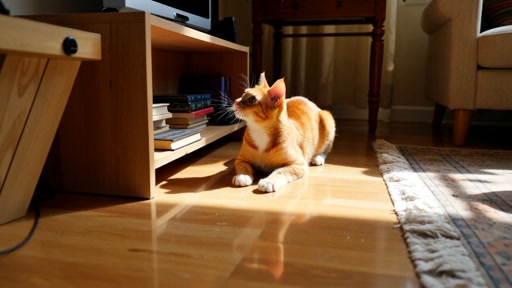}%
\includegraphics[width=0.333\linewidth]{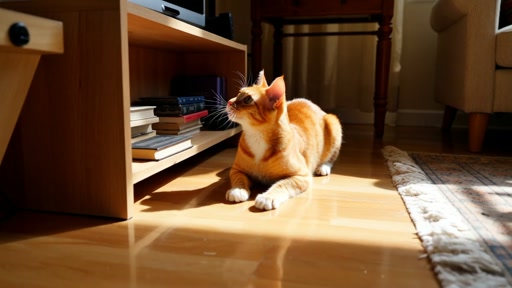}%
\includegraphics[width=0.333\linewidth]{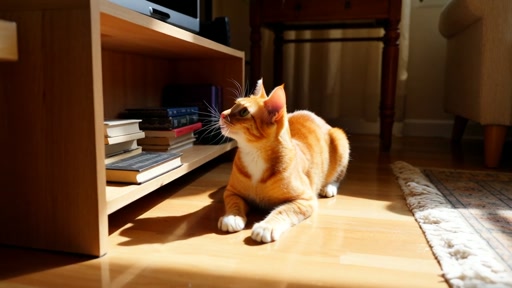}}%
} &
\vbox{
\hbox{\animategraphics[width=\linewidth]{16}{videos/without_cache/}{0000}{0080}}
\hbox{\animategraphics[width=\linewidth]{16}{videos/with_cache_10step/}{0000}{0080}}
\hbox{\animategraphics[width=\linewidth]{16}{videos/ref/}{0000}{0080}}
\vspace{5pt}
\hbox{\animategraphics[width=\linewidth]{16}{videos/without_cache_1/}{0000}{0080}}
\hbox{\animategraphics[width=\linewidth]{16}{videos/with_cache_10step_1/}{0000}{0080}}
\hbox{\animategraphics[width=\linewidth]{16}{videos/ref1/}{0000}{0080}}
} \\
\end{tabular}
}
\vspace{-5mm}
\caption{Two video generation examples, each with three rows. ``w/o cache'' is generated directly from the prompt, ``w/ cache'' reuses latents from a similar prompt, and ``ref'' shows the generation from the similar prompt. Despite high similarity scores, both results exhibit obvious semantic drift.}
\label{fig:teaser}
\end{figure}

\noindent\textbf{Discussion}. Using cached latents can introduce approximation errors (reconstruction or distributional mismatch) and thereby degrade final video quality. To implement effective scaling, we posit that an effective caching strategy is crucial, and several challenges must be addressed properly.

\begin{itemize}
    \item\textit{Similarity Measurement}. Determining similarity in a caching strategy is essential, as it directly affects both the cache hit rate and the quality of the generated videos. While existing approaches commonly rely on CLIP embeddings to measure similarity, we observe that reusing cached results from a highly ``similar'' request does not always guarantee satisfactory video quality. In particular, semantic similarity at the text or image level may fail to capture fine-grained temporal dynamics or motion patterns that are crucial for video generation. We illustrate two representative failure cases in~\autoref{fig:teaser}.
    
    \item\textit{Reuse Criteria}. To ensure video quality while achieving computational savings, it is equally important to determine appropriate reuse depths under varying similarity scores. Reusing latents only at shallow depth generally preserves visual fidelity but yields limited performance gains, whereas aggressive reuse at deeper steps can significantly reduce computation at the cost of noticeable quality degradation when similarity is imperfect. This trade-off highlights the need for adaptive reuse criteria that jointly consider similarity confidence and diffusion depth.
\end{itemize}

\subsection{Multi-modal Data Pre-Processing with Spatial Sharing}\label{sec:4.2}
\subsubsection{Background}
As ``\textit{garbage in, garbage out}'' implies, the quality of training data determines model performance~\citep{data-quality20, phi-1-23}. To ensure high-quality data inputs, existing models rely on sophisticated data pre-processing pipelines to curate data from vast raw sources. In contrast to traditional machine learning (ML) input data pipelines that were integrated into the training job and ran on CPUs~\citep{Cachew22, GoldMiner23, FastFlow23}, contemporary multi-modal data processing pipelines have grown increasingly complex and GPU-intensive, and are often decoupled from model training~\citep{Open-Sora24, wan25, hunyuan3d22025tencent, Seedance25}. \autoref{fig:open-sora-pipeline} depicts a representative data curation pipeline adopted in training T2V models~\citep{Open-Sora24}.

% \definecolor{bluebox}{RGB}{133,151,191}
% \begin{wrapfigure}{r}{0.55\textwidth}
%     \vspace{-10pt}
%     \begin{center}
%         \includegraphics[width=\linewidth]{figures/pipeline.pdf}
%     \end{center}
%     \vspace{-1em}
%     \caption{A representative processing pipeline where \textcolor{graybox}{\textbf{Gray boxes}} are CPU-bound and \textcolor{bluebox}{\textbf{blue boxes}} are GPU-bound operations.}
%     \label{fig:open-sora-pipeline}
% \end{wrapfigure}
\definecolor{bluebox}{RGB}{133,151,191}
\begin{figure}[h]
    \centering
    \vspace{-2mm}
    \includegraphics[width=0.8\linewidth]{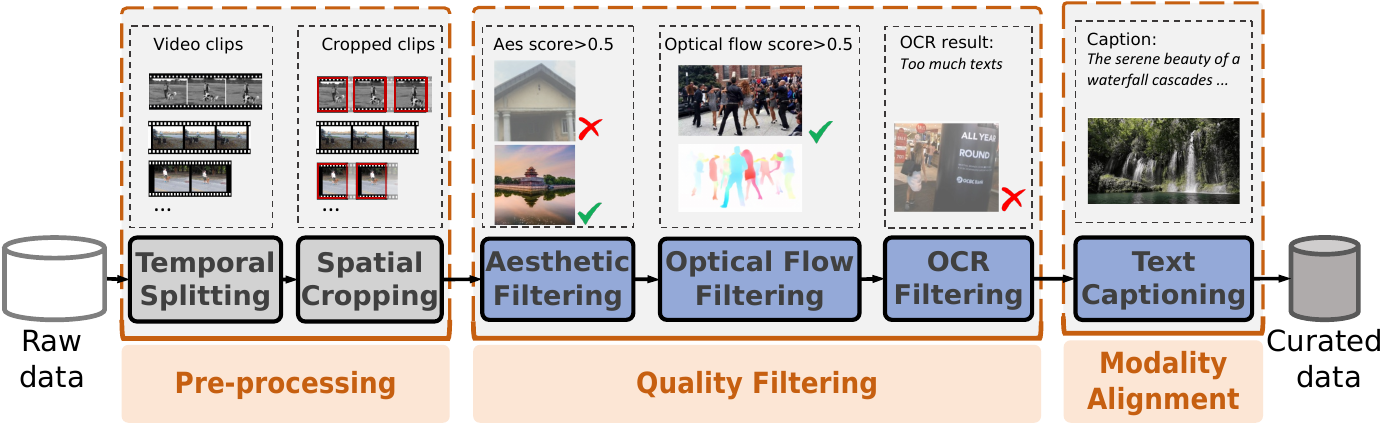}
    % \vspace{-2mm}
    \caption{A representative data processing pipeline where \textcolor{graybox}{\textbf{Gray boxes}} are CPU-bound and \textcolor{bluebox}{\textbf{blue boxes}} are GPU-bound operations.}
    \label{fig:open-sora-pipeline}
\end{figure}

\begin{wrapfigure}{r}{0.55\textwidth}
    % \begin{center}
    \vspace{2mm}
    \includegraphics[width=\linewidth]{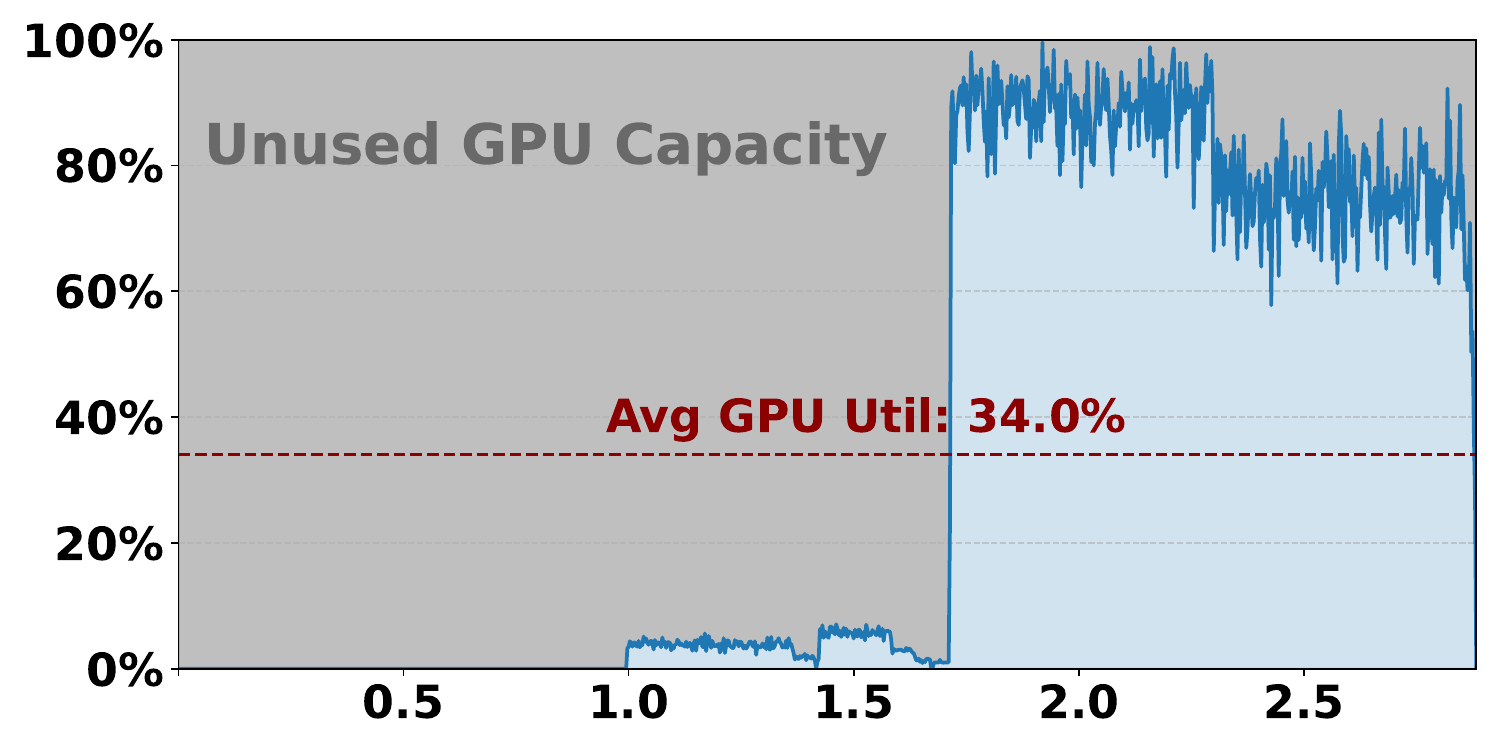}
    \vspace{-7mm}
    \caption{An observation of low GPU utilization (\%) in the existing data pre-processing procedure.}
    % \end{center}
    \label{fig:gpu-util}
\end{wrapfigure}

\noindent\textbf{Challenges}. However, the complexity of curating multimodal data with such pipelines incurs substantially greater computational overhead than traditional workflows, and often results in poor resource utilization, particularly of GPUs. For example, we assess the performance of one of the state-of-the-art processing frameworks~\citep{data-juicer2024} on video data from a public dataset using the curation pipeline (right). \autoref{fig:gpu-util} plots the GPU utilization over the entire processing pipeline. We observe that at the start of the curation, GPU-bound operators (OPs) cannot commence until CPU-bound OPs complete, leaving GPUs idle for the first hour; Thereafter, although multiple heterogeneous models are continuously assigned to workers via the underlying distributed execution engine, GPU resources are not fully saturated for most of the runtime, resulting in an average GPU utilization of only $34.0$\%. Thus, finding an effective way to reduce underutilized computational resources is essential for maximizing the overall performance of data processing systems.

\noindent\textbf{Proposed Method}. Motivated by the architectures of ML inference serving systems~\citep{triton24}, we re-conceptualize multi-modal data pre-processing as a multi-model inference serving problem. To mitigate the computational inefficiency arising from heterogeneous GPU workloads, we propose leveraging \textit{spatial sharing techniques}~\citep{mps}, which enable multiple models to concurrently reside on and execute on the same GPU. This design explicitly exploits the trade-off between GPU memory and compute efficiency, converting otherwise idle GPU resources into productive computation by allocating additional memory for multi-model co-residency.

\subsubsection{Preliminary Results}
\definecolor{dscolor}{RGB}{27,158,119} % teal green 
\definecolor{mlmcolor}{RGB}{217,95,2} % burnt orange 
\definecolor{fspcolor}{rgb}{0.4,0.0,0.6} % magenta (MuMoMo Ray)
\begin{figure}
\hspace{-.3cm}
\begin{tikzpicture}
\begin{groupplot}[
    group style={
        group size=4 by 2, % <<< 改成两行四列
        horizontal sep=0.6cm,
        vertical sep=1.cm  % <<< 控制行间间距
    },
    width=5.cm,
    height=4.cm,
    xmin=0, 
    xmax=100, 
    ymin=0,
    ymajorgrids=true,
    enlarge x limits=0.05,
    enlarge y limits=0.05,
    title style={yshift=-1.2ex,font=\scriptsize},
    legend style={
        at={(-1.27,2.85)}, % <<< 适当调高以适应两行
        anchor=north,
        legend columns=3,
        line width=0.1pt,
        font=\scriptsize
    }
]
% --- 第一行 4 个子图 ---
\nextgroupplot[title={Pipeline=T2V, Dataset=OpenVid}, xtick={0,20,50,70,100}, xticklabel style={font=\small}, ymax=25, ytick={0,4,10,20,25}, ylabel={time (hrs)}, ylabel style={yshift=-2ex,font=\scriptsize}]
\addplot[mark=*, mark size=1.5, line width=1.5pt, color=dscolor] coordinates {(0,0) (20,5.12) (50,5.66) (70,5.92) (100,6.05)};%(5,4.93)
\addplot[mark=square*, mark size=1.5, line width=1.5pt, color=mlmcolor] coordinates {(0,0) (20,4.22) (50,10.99) (70,15.65) (100,22.63)};%(5,0.97)
\addplot[mark=diamond*, mark size=2.2, line width=1.5pt, color=fspcolor, mark options={scale=1, xscale=1.2, yscale=1}] coordinates {(0,0) (20,1.61) (50,2.07) (70,2.31) (100,2.78)};%(5,1.24)

\nextgroupplot[title={Pipeline=T2V, Dataset=Koala}, xtick={0,20,50,70,100}, xticklabel style={font=\small}, ymax=25, ylabel={}, ytick={0,4,10,20,25}]
\addplot[mark=*, line width=1.5pt, color=dscolor] coordinates {(0,0) (20,5.50) (50,6.83) (70,7.52) (100,8.20)};%(5,4.95)
\addplot[mark=square*, line width=1.5pt, color=mlmcolor] coordinates {(0,0) (20,2.99) (50,6.03) (70,15.11) (100,21.29)};% (5,1.45)
\addplot[mark=diamond*, line width=1.5pt, color=fspcolor, mark size=2.2, mark options={scale=1, xscale=1.2, yscale=1}] coordinates {(0,0) (20,2.07) (50,2.70) (70,3.15) (100,3.58)};%(5,1.57) 

\nextgroupplot[title={Pipeline=T2I, Dataset=LAION}, xtick={0,20,50,70,100}, xticklabel style={font=\small}, ylabel={}, ymax=9, ytick={0,3,6,9}]
\addplot[mark=*, line width=1.5pt, color=dscolor] coordinates {(0,0) (20,0.71) (50,1.78) (70,2.60) (100,3.70)};%(5,0.16) 
\addplot[mark=square*, line width=1.5pt, color=mlmcolor] coordinates {(0,0) (20,1.67) (50,4.21) (70,5.91) (100,8.48)};
\addplot[mark=diamond*, line width=1.5pt, color=fspcolor, mark size=2.2, mark options={scale=1, xscale=1.2, yscale=1}] coordinates {(0,0) (20,0.41) (50,1.02) (70,1.43) (100,2.05)};%(5, 0.10)

\nextgroupplot[title={Pipeline=T2I, Dataset=COYO}, xtick={0,20,50,70,100}, xticklabel style={font=\small}, ylabel={}, ymax=1.8, ytick={0,0.5,1.0,1.5,1.8}]
\addplot[mark=*, line width=1.5pt, color=dscolor] coordinates {(0,0) (5,0.13) (20,0.39) (50,0.65) (70,1.02) (100,1.50)};
\addplot[mark=square*, line width=1.5pt, color=mlmcolor] coordinates {(0,0) (5,0.10) (20,0.41) (50,0.70) (70,1.10) (100,1.56)};
\addplot[mark=diamond*, line width=1.5pt, color=fspcolor, mark size=2.2, mark options={scale=1, xscale=1.2, yscale=1}] coordinates {(0,0) (5, 0.06) (20,0.24) (50,0.59) (70,0.83) (100,1.19)};

% --- 第二行 4 个子图 ---
\nextgroupplot[
    xtick={5,20,50,70,100}, 
    xlabel={num of data samples (k)},
    xlabel style={
        at={(axis description cs:0.5,0.04)},
        anchor=north,
        font=\scriptsize
    }, 
    ylabel={throughput (samples/s)}, 
    ylabel style={yshift=-2ex,font=\scriptsize},
    ymax=12,
    ytick={0,4,8,12}
]
\addplot[mark=*, line width=1.5pt, color=dscolor] coordinates {(5,0.28) (20,1.09) (50,2.45) (70,3.24) (100,4.59)};
\addplot[mark=square*, line width=1.5pt, color=mlmcolor] coordinates {(5,1.43) (20,1.32) (50,1.26) (70,1.24) (100,1.23)};
\addplot[mark=diamond*, line width=1.5pt, color=fspcolor, mark size=2.2, mark options={scale=1, xscale=1.2, yscale=1}] coordinates {(5,1.12) (20,3.45) (50,6.71) (70,8.42) (100,9.99)};

\nextgroupplot[
    xtick={5,20,50,70,100}, 
    ymax=9,
    ylabel={}, ytick={0,3,6,9},
    yticklabels={0,3,6,9},
    xlabel={num of data samples (k)},
    xlabel style={
        at={(axis description cs:0.5,0.04)},
        anchor=north,
        font=\scriptsize
    }
]
\addplot[mark=*, line width=1.5pt, color=dscolor] coordinates {(5,0.28) (20,1.01) (50,2.03) (70,2.59) (100,3.39)};
\addplot[mark=square*, line width=1.5pt, color=mlmcolor] coordinates {(5,0.96) (20,1.86) (50,2.30) (70,1.27) (100,1.30)};
\addplot[mark=diamond*, line width=1.5pt, color=fspcolor, mark size=2.2, mark options={scale=1, xscale=1.2, yscale=1}] coordinates {(5,0.88) (20,2.68) (50,5.14) (70,6.17) (100,7.76)};

\nextgroupplot[
    xtick={5,20,50,70,100}, 
    ylabel={}, ymax=15,
    ytick={0,5,10,15},
    yticklabels={0,5,10,15},
    xlabel={num of data samples (k)},
    xlabel style={
        at={(axis description cs:0.5,0.04)},
        anchor=north,
        font=\scriptsize
    }
]
\addplot[mark=*, line width=1.5pt, color=dscolor] coordinates {(5,8.68) (20,7.82) (50,7.80) (70,7.48) (100,7.51)};
\addplot[mark=square*, line width=1.5pt, color=mlmcolor] coordinates {(5,3.23) (20,3.33) (50,3.30) (70,3.29) (100,3.28)};
\addplot[mark=diamond*, line width=1.5pt, color=fspcolor, mark size=2.2, mark options={scale=1, xscale=1.2, yscale=1}] coordinates {(5,13.89)(20,13.55) (50,13.62) (70,13.60) (100,13.55)};

\nextgroupplot[
    xtick={5,20,50,70,100}, 
    ylabel={},ymax=25,
    ytick={0,10,20,25},
    yticklabels={0,10,20,25},
    xlabel={num of data samples (k)},
    xlabel style={
        at={(axis description cs:0.5,0.04)},
        anchor=north,
        font=\scriptsize
    }
]
\addplot[mark=*, line width=1.5pt, color=dscolor] coordinates {(5,10.68) (20,14.25) (50,21.37) (70,19.06) (100,18.52)};
\addplot[mark=square*, line width=1.5pt, color=mlmcolor] coordinates {(5,13.89) (20,13.55) (50,19.84) (70,17.68) (100,17.81)};
\addplot[mark=diamond*, line width=1.5pt, color=fspcolor, mark size=2.2, mark options={scale=1, xscale=1.2, yscale=1}] coordinates {(5,23.15) (20,23.15) (50,23.54) (70,23.43) (100,23.34)};

\legend{Data-Juicer,Daft,(ours)}

\end{groupplot}
\end{tikzpicture}
\caption{Performance of processing time (top row) and throughput (bottom row) across different settings. Subcaption of each group (column) indicates the corresponding pipeline and dataset.}
\label{fig:overall-results}
\end{figure}
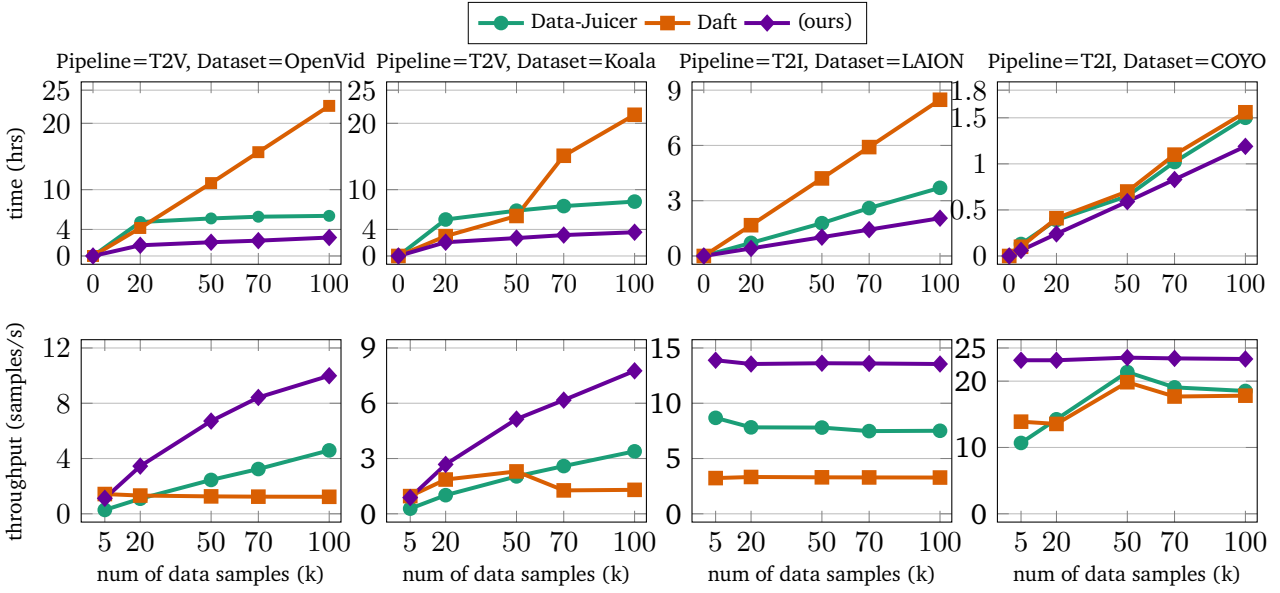

\begin{wrapfigure}{r}{.5\textwidth}
    \vspace{-3mm}
    \begin{subfigure}[b]{\linewidth}
        \centering
        \includegraphics[width=\linewidth]{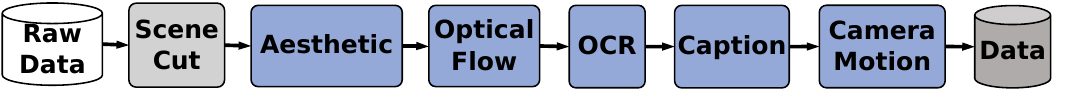}
        \caption{T2V}
        \label{fig:t2v-pipe}
    \end{subfigure}
    \bigskip
    \begin{subfigure}[b]{\linewidth}
        \centering
        \includegraphics[width=\linewidth]{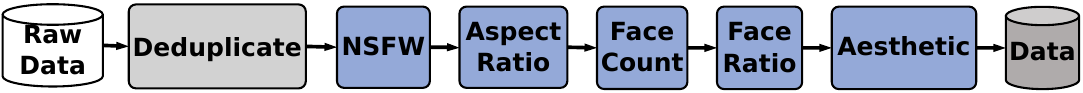}
        \caption{T2I}
        \label{fig:t2i-pipe}
    \end{subfigure}
    \vspace{-13mm}
\caption{T2V and T2I processing pipelines.}
\label{fig:pipelines}
\end{wrapfigure}

To evaluate data pre-processing performance under realistic conditions, we use two modern multi-modal data curation pipelines employed in practical T2V and T2I scenarios, as shown in~\autoref{fig:pipelines}.

% \begin{figure}[h]
%     \vspace{5mm}
%     \begin{subfigure}[b]{0.8\linewidth}
%         \centering
%         \includegraphics[width=0.8\linewidth]{figures/t2v-pipe.pdf}
%         \caption{Text-to-video (T2V) processing pipeline}
%         \label{fig:t2v-pipe}
%     \end{subfigure}
%     \bigskip
%     \begin{subfigure}[b]{0.8\linewidth}
%         \centering
%         \includegraphics[width=0.8\linewidth]{figures/t2i-pipe.pdf}
%         \caption{Text-to-image (T2I) processing pipeline}
%         \label{fig:t2i-pipe}
%     \end{subfigure}
% \caption{Typical multi-modal data curation pipelines employed in practical T2V and T2I scenarios.}
% \label{fig:pipelines}
% \end{figure}

\noindent\textbf{Datasets}. We evaluate on four public multi-modal datasets: OpenVid~\citep{OpenVid25} and Koala~\citep{Koala25} used for the T2V training, as well as COYO~\citep{COYO22} and LAION~\citep{LAION22} for the T2I training.

\noindent\textbf{Baselines}. We utilize two state-of-the-art data pre-processing systems, namely Data-Juicer~\citep{data-juicer2024} and Daft~\citep{daft25}, as baselines for comparison.

\noindent\textbf{Metrics}. We select to use four commonly used performance metrics for the evaluation: \textit{processing time}, \textit{throughput}, \textit{average CPU consumption}, and \textit{GPU utilization}.

\noindent\textbf{Results}. The overall results are presented in~\autoref{fig:overall-results}. The framework demonstrates exceptional efficiency on video data, achieving up to a $2.29\times$ reduction in processing latency on the Koala dataset compared to Data-Juicer (second figure in the top row), and reaching a throughput of up to $10$ video samples per second on the OpenVid dataset (first figure in the bottom row). Furthermore, although all systems exhibit better processing performance on text-image data than on text-video data, performance varies with dataset characteristics.

\definecolor{tropicalrainforest}{rgb}{0.66, 0.82, 0.56}
\definecolor{trolleygrey}{rgb}{0.5, 0.5, 0.49}
\definecolor{iryellow}{rgb}{1.0, 0.91, 0.64}
\definecolor{glaucous}{rgb}{0.38, 0.51, 0.71}
\definecolor{cardinal}{rgb}{0.77, 0.33, 0.3}
\definecolor{orange}{rgb}{0.96, 0.69, 0.51}
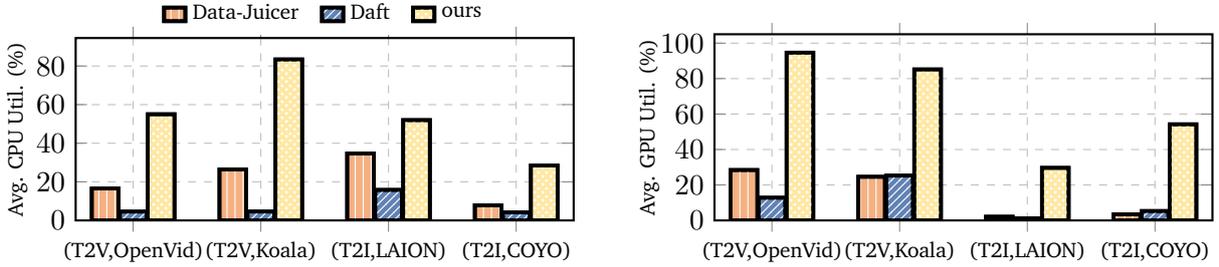
\begin{figure}[h]
    \centering
    \begin{subfigure}[b]{0.49\linewidth}
        \centering
        \begin{tikzpicture}
        \begin{axis}[
            height=4.cm,
            width=\linewidth,
            ybar = .17mm, 
            enlarge y limits=0,
            enlarge x limits=0.15, 
            ymin = 0,
            ymax = 90,
            ytick={0,20,40,60,80},
            axis line style=thick,
            ylabel={Avg. CPU Util. (\%)},
            ylabel style={yshift=-2.5ex,font=\scriptsize},
            symbolic x coords={{(T2V,OpenVid)}, {(T2V,Koala)}, {(T2I,LAION)}, {(T2I,COYO)}}, 
            xtick=data,
            xticklabel style={
                font=\scriptsize,
            },
            legend image code/.code={%
                \draw[#1, draw=black] (0cm,-0.1cm) rectangle (0.28cm,0.08cm);
            },
            bar width=10pt,
            enlarge y limits={0.05, upper},
            xmajorgrids=true,
            ymajorgrids=true,
            grid style=dashed,
            nodes near coords align={vertical}, 
            legend pos=north east,
            legend style={
                font=\scriptsize,
                at={(0.5,1.25)},
                legend columns=-1,
                anchor=north,
                /tikz/column 2/.style={column sep=6pt},
                /tikz/column 4/.style={column sep=6pt},
                draw=none
            }
        ]
        \addplot [
            line width= .5mm, 
            fill=orange,
            postaction={
                pattern=vertical lines, pattern color=white
            }
        ]
        coordinates {({(T2V,OpenVid)},16.53) ({(T2V,Koala)},26.42) ({(T2I,LAION)},34.68) ({(T2I,COYO)},7.85)};
        \addplot[
            line width= .5mm, 
            fill=glaucous,
            postaction={
                pattern=north east lines, pattern color=white
            }
        ]
        coordinates {({(T2V,OpenVid)},4.58) ({(T2V,Koala)},4.61) ({(T2I,LAION)},15.82) ({(T2I,COYO)},4.16)};
        \addplot [
            line width= .5mm,
            fill=iryellow,
            postaction={
                pattern=crosshatch dots, pattern color=white
            }
        ]
        coordinates {({(T2V,OpenVid)},55.02) ({(T2V,Koala)},83.43) ({(T2I,LAION)},51.98) ({(T2I,COYO)},28.42)};
        \legend{Data-Juicer, Daft, ours}
        \end{axis}
        \end{tikzpicture}
        \vspace{-1mm}
        \caption{Average CPU consumption.}
        \label{fig:cpu-util-results}
    \end{subfigure}
    \hfill
    \begin{subfigure}[b]{0.49\linewidth}
        \centering
        \begin{tikzpicture}
        \begin{axis}[
            height=4.05cm,
            width=\linewidth,
            ybar = .17mm, 
            enlarge y limits=0,
            enlarge x limits=0.15, 
            ymin = 0,
            ymax = 100,
            ytick={0,20,40,60,80,100},
            axis line style=thick,
            ylabel={Avg. GPU Util. (\%)},
            ylabel style={yshift=-2ex,font=\scriptsize},
            symbolic x coords={{(T2V,OpenVid)}, {(T2V,Koala)}, {(T2I,LAION)}, {(T2I,COYO)}}, 
            xtick=data,
            xticklabel style={font=\scriptsize},
            bar width=10pt,
            enlarge y limits={0.05, upper},
            xmajorgrids=true,
            ymajorgrids=true,
            grid style=dashed,
            nodes near coords align={vertical}, 
        ]
        \addplot [
            line width= .5mm, 
            fill=orange,
            postaction={
                pattern=vertical lines, pattern color=white
            }
        ]
        coordinates {({(T2V,OpenVid)},28.43) ({(T2V,Koala)},24.62) ({(T2I,LAION)},2.10) ({(T2I,COYO)},3.46)};
        \addplot[
            line width= .5mm, 
            fill=glaucous,
            postaction={
                pattern=north east lines, pattern color=white
            }
        ]
        coordinates {({(T2V,OpenVid)},12.82) ({(T2V,Koala)},25.28) ({(T2I,LAION)},1.19) ({(T2I,COYO)},5.30)};
        \addplot [
            line width= .5mm,
            fill=iryellow,
            postaction={
                pattern=crosshatch dots, pattern color=white
            }
        ]
        coordinates {({(T2V,OpenVid)}, 94.64) ({(T2V,Koala)},85.19) ({(T2I,LAION)},29.65) ({(T2I,COYO)},54.22)};
        \end{axis}
        \end{tikzpicture}
        \vspace{-1mm}
        \caption{Average GPU utilization.}
        \label{fig:gpu-util-results}
    \end{subfigure}
    \vspace{-2mm}
    \caption{Performance of average CPU and GPU utilization.}
    \label{fig:cpu-gpu-util-results}
\end{figure}

On the other hand, our framework achieves substantially higher hardware utilization (both CPU and GPU) than the two baseline systems, as evident in~\autoref{fig:cpu-gpu-util-results}. In terms of CPU consumption (\autoref{fig:cpu-util-results}), the framework attains an average CPU utilization of $69.22$\% across the two T2V datasets, despite not employing any specific optimizations for CPU resource usage. In terms of GPU utilization (\autoref{fig:gpu-util-results}), the framework shows a substantial improvement over the two baselines, largely because both baselines exhibit poor GPU performance. One notable outcome is observed when applying the T2V pipeline to OpenVid: our framework achieves an average GPU utilization of $94.64\%$, whereas Data-Juicer and Daft reach only $28.43\%$ and $12.82\%$, respectively. These results further demonstrate the effectiveness of trading increased GPU memory usage for reduced unnecessary computation in modern multimodal data pre-processing scenarios.

\subsection{Summary}
This section demonstrates that memory can be effectively traded for reduced computation in modern AI systems, providing a practical optimization path under compute-constrained settings. Through two diverse scenarios\textemdash diffusion-based video inference (\S\ref{sec:4.1}) and multimodal data pre-processing (\S\ref{sec:4.2})\textemdash we show how allocating additional storage (including CPU and GPU memory) at different system layers can systematically reduce computational overhead.

The key takeaway from these use cases for practitioners is that workloads exhibiting redundancy or low resource utilization benefit substantially from memory-first system designs. As memory continues to scale faster and cheaper than compute, \textbf{deliberately leveraging memory capacity to trade off prohibitive computation should be a core design principle for building next-generation, large-scale AI systems}.

% Together, these examples illustrate a unifying principle: when bandwidth is scarce, computation can act as a powerful substitute for communication. By exploiting learned priors, semantic representations, and task structure, AI systems can move beyond transmitting raw data and instead communicate only what is essential. This trade-off provides a flexible design paradigm for deploying intelligent services in bandwidth-constrained environments, enabling efficient system optimization
\section{Future Work}\label{sec:6}
 The AI Trinity framework offers a novel approach to dynamic resource balancing, though multiple avenues remain for further investigation. Below, we provide three forward-looking directions to mature the AI Trinity framework and extend its impact across the modern AI stack.
 
\subsection{Cycling AI Trinity}
The current AI Trinity framework typically involves unidirectional resource exchanges between computation, memory, and bandwidth. However, exploring a more advanced \textit{Cycling AI Trinity} in which resources cycle in both directions remains unexplored. 

For instance, in edge devices with constrained processing power, computationally-intensive tasks can be offloaded to the cloud, enabling efficient processing without overburdening the local device. This bandwidth$\rightarrow$computation trade-off can be beneficial in bandwidth-rich environments like 5G networks, where the primary constraint is not bandwidth but local computational capacity. By increasing bandwidth utilization and shifting the computation to more powerful remote systems, the system can maintain high throughput without incurring significant delays. Another use case of the potential computation$\rightarrow$memory trade-off is in resource-constrained devices, where memory capacity is limited, such as mobile devices or IoT sensors. In such environments, pre-processing or compressing data locally via additional computation before storage or transmission can alleviate memory constraints.

% Hence, a fully closed-loop framework, where $computation \leftrightarrow bandwidth \leftrightarrow memory$, would enable a more fluid and scalable system that can adjust to the demands of diverse AI tasks across edge, cloud, and hybrid environments.
Hence, a fully closed-loop AI Trinity framework, \textit{i.e.}, \textbf{\textcolor{comp}{Computation}}$\leftrightarrow$\textbf{\textcolor{band}{Bandwidth}}$\leftrightarrow$\textbf{\textcolor{mem}{Memory}}, would enable a more fluid and scalable system that can adjust to the demands of diverse AI tasks across edge, cloud, and hybrid environments.

\subsection{Green AI Trinity}
As AI systems grow more complex and energy-intensive, their environmental impact is coming under increasing scrutiny. An encouraging avenue for future research is adapting the AI Trinity framework to prioritize energy efficiency. Green AI Trinity aims to minimize the ecological footprint of AI systems by making energy consumption a key factor in resource-allocation decisions, without compromising system performance. By incorporating sustainability into system design, Green AI Trinity helps ensure that AI advancements do not come at the expense of the planet's future.

The core idea behind Green AI Trinity is to extend the traditional resource trade-offs to include energy consumption as an essential variable. Much like AI Trinity's ability to dynamically balance computation and memory resources to alleviate bottlenecks, Green AI Trinity introduces energy-aware resource management. For example, it would involve strategies such as shifting more computation to energy-efficient edge devices or reducing the energy cost of data transmission through optimized bandwidth usage. By incorporating power-efficient algorithms and hardware components, Green AI Trinity enables AI systems to achieve near-optimal performance while minimizing energy consumption and, by extension, their environmental impact.

Achieving the Green AI Trinity will require future research to develop innovative methods for incorporating energy-consumption models into AI systems' decision-making processes. This includes developing energy-efficient AI algorithms, improving the design of hardware accelerators for AI workloads, and optimizing software frameworks to minimize power use. Additionally, AI models themselves could be optimized for energy efficiency without sacrificing accuracy or capabilities. By making energy efficiency an integral part of the AI system's design, Green AI Trinity offers a pathway towards more sustainable AI development, harmonizing technological advancement with global initiatives to fight climate change and lower carbon emissions in the tech sector.

\section{Conclusion}\label{sec:7}
In this paper, we presented Computation-Bandwidth-Memory Trade-offs, or AI Trinity, a novel design principle for next-generation AI systems. The core idea of AI Trinity is to elevate computation, memory, and bandwidth to coequal pillars for modern AI infrastructure, and to achieve system-level performance optimality by dynamically balancing these resources to meet specific application requirements. AI Trinity formulated a closed-loop set of resource-exchange pathways and identified three fundamental trade-offs: trading computation for bandwidth in network-constrained scenarios, trading bandwidth for memory in memory-limited scenarios, and trading memory for computation in compute-intensive scenarios. We demonstrated the applicability of AI Trinity through a range of representative AI systems spanning the full AI development lifecycle, illustrating how AI Trinity translates the principle of resource balancing into concrete design choices. Finally, we outlined directions for future research to expand the scope of AI Trinity. Overall, this paper presents a unified paradigm for next-generation AI infrastructure design, empowering practitioners to rethink system design decisions beyond naive resource scaling.

\clearpage

\bibliography{main}
\bibliographystyle{iclr2025_conference}

\appendix

\end{document}